\RequirePackage{fix-cm}
\documentclass{svjour3}                     

\smartqed  
\usepackage{amssymb}
\usepackage{comment}
\usepackage{graphicx}
\usepackage{subcaption}
\usepackage{hyperref}
\usepackage{booktabs}   
\usepackage{multirow}   
\usepackage{lmodern}    
\usepackage{siunitx}
\usepackage{etoolbox}

\makeatletter
\patchcmd{\@maketitle}{\vskip7.23pt}{\vskip2pt}{}{}

\patchcmd{\@maketitle}{\vskip 7.2mm}{\vskip 2mm}{}{}
\patchcmd{\@maketitle}{\vskip 5.2mm}{\vskip 1.5mm}{}{}
\makeatother

\journalname{Empirical Software Engineering}

\begin{document}

\title{Validation of an analyzability model for quantum software: a family of experiments}

\titlerunning{Validation of an analyzability model for quantum software}        

\author{Ana Díaz-Muñoz \and José A. Cruz-Lemus \and Moisés Rodríguez 
\and Maria Teresa Baldassarre \and Mario Piattini
}


\institute{Ana Díaz-Muñoz \at
              AQCLab Software Quality and University of Castilla-La Mancha \\
              \email{adiaz@aqclab.es}           
           \and
           José A. Cruz-Lemus \at
              University of Castilla-La Mancha \\
              \email{joseantonio.cruz@uclm.es}
           \and
           Moisés Rodríguez \at
              AQCLab Software Quality and University of Castilla-La Mancha \\
              \email{moises.rodriguez@uclm.es}
           \and
           Maria Teresa Baldassarre \at
              University of Bari Aldo Moro \\
              \email{mariateresa.baldassarre@uniba.it}
           \and
           Mario Piattini \at
              University of Castilla-La Mancha \\
              \email{mario.piattini@uclm.es}          
}

\date{Received: 30 May 2025 / Accepted: 1 February 2026}

\maketitle

\noindent\textbf{Note:} This is the author accepted manuscript of the article published in
\emph{Empirical Software Engineering} (2026). The final published version is available at
\url{https://doi.org/10.1007/s10664-026-10825-3}.

\begin{abstract}
The analyzability of hybrid software, which integrates both classical and quantum components, is a key factor in ensuring its maintainability and industrial adoption. This article presents the empirical validation, through a family of experiments, of the quantum component of a previously proposed hybrid software analyzability model based on the ISO/IEC 25010 standard. The experimental series consists of four studies involving participants with diverse profiles in both academic and professional settings. In these experiments, the model's ability to effectively measure the analyzability of quantum algorithms is assessed, and the relationship between the analyzability levels computed by the model and the participant’s perceptions of the complexity of these algorithms is examined. The results indicate that the proposed model effectively distinguishes between quantum software components with varying levels of analyzability and aligns with human perception, reinforcing its validity in quantum computing.



\end{abstract}

\section{Introduction}
\label{intro}

The development of quantum software, which exploits the unique processing capabilities inherent to quantum mechanics, has gained increasing importance in quantum computing and software engineering. These systems harness quantum phenomena to solve problems that remain intractable for classical computers \cite{bernardt2019quantum} \cite{junior2024systematicmappingquantumsoftware} \cite{piattini2021quantum}, such as combinatorial optimization, complex physical system simulations, and advanced cryptography. However, due to the inherent challenges of quantum technology, fully quantum systems are currently not viable for most applications. As a consequence, in practice, quantum software is developed by integrating quantum algorithms with supportive classical components. This integration allows for an accelerated processing power and efficiency of quantum algorithms to be utilized alongside the robustness and reliability of classical systems, resulting in hybrid software that maximizes the benefits of quantum processing.

The rise of quantum computing has driven the development of architectures focused on enhancing quantum capabilities to address complex problems beyond the reach of conventional systems. Landmark experiments—such as the quantum supremacy demonstration \cite{arute2019quantumsupremacy}—have illustrated the potential of quantum algorithms for solving specific problems faster than classical systems; however, the scalability and reliability of purely quantum systems still face technological limitations, reinforcing the need to adopt approaches that fully leverage the theoretical advantages of quantum processing while mitigating current hardware constraints \cite{preskill2018quantumcomputing}.

Despite its promise, quantum software development presents unique challenges arising from the fundamentally probabilistic nature of quantum computation. While classical software relies on deterministic operations over well-defined data structures, quantum circuits operate probabilistically and utilize qubits in superposition and entanglement, complicating analysis and debugging \cite{AKTAR2025107587}. Additionally, the no-cloning theorem and the need to measure qubits to extract information impose further constraints on the development and maintenance of quantum systems \cite{zappin2024quantummeetsclassicalcharacterizing}. Therefore, maintaining the quality of quantum software demands both the adaptation of existing metrics and the creation of new methodologies tailored to assess its unique quantum characteristics.

Software quality is a key factor in ensuring the reliability, efficiency, and sustainability of quantum systems \cite{Colakoglu2021} \cite{piattini2015iso25000}. The ISO/IEC 25010 standard \cite{iso25010} provides a reference framework for classical software products, identifying essential characteristics such as maintainability, usability, and reliability \cite{rodriguez2016functional}. However, when applied to quantum software, these models require significant adaptation since the unique characteristics of quantum computing demand additional quality indicators. One of the most critical sub-characteristics in this context is \textit{analyzability}, defined by ISO/IEC 25010 as ``the degree of effectiveness and efficiency with which it is possible to assess the impact on a product or system of an intended change, to diagnose a product for deficiencies or causes of failures, or to identify parts to be modified'' \cite{iso25010}. In quantum systems, analyzability is particularly critical because developers must integrate classical computational approaches with the novel principles of quantum algorithms.

While well-established quality models exist for classical software \cite{piattini2020talavera} \cite{Verdugo2024Laboratory}, their direct application to quantum software is quite insufficient, as they do not account for the specificities of quantum computing \cite{gao2021quantum}. Recent research \cite{diaz2024environment} \cite{diaz2024metrics} has begun to address this gap by incorporating classical and quantum metrics into a unified evaluation framework, integrating indicators such as adapted cyclomatic complexity and quantum circuit depth. Despite these advances, there is no consolidated model that can comprehensively assess analyzability in quantum environments \cite{liu2022benchmarking}. Previous studies \cite{diaz2024environment} \cite{diaz2024metrics}\cite{diaz2025environment} have proposed a quality model based on ISO/IEC 25010 that combines traditional metrics with novel quantum-specific indicators to evaluate quantum circuits. However, validating these quantum metrics requires experiments that incorporate diverse subject profiles and experimental conditions.

This article presents a family of experiments that aim to empirically validate the analyzability model for quantum software. The study was originally proposed as a Registered Report, accepted in the 18th ACM/IEEE International Symposium on Empirical Software Engineering (ESEM 2024) Registered Reports track 
\cite{ana2024validationanalysabilitymodelhybrid}. The original features and design of the family of experiments (Stage 1) have been followed. 
Slight adjustments are commented in section \ref{sub-sec:materials}. Each experiment assesses the analyzability of quantum circuits in different contexts and examines the relationship between the analyzability levels computed by the model and the participants’ perceptions of complexity. Each experiment assesses the analyzability of quantum circuits in different contexts and examines the relationship between the analyzability levels computed by the model and the participants’ perceptions of complexity. Furthermore, evidence is provided on the necessity of focusing on quantum-specific quality evaluation, as leveraging the unique properties of quantum processing is essential for achieving optimal performance. Various studies have demonstrated the effectiveness of variational approaches and machine learning techniques in optimizing quantum circuits and mitigating errors in current hardware devices \cite{biamonte2017quantum} \cite{endo2021practical} \cite{kandala2017hardware}. These advancements furnish new analytical tools that can be adapted to assess the quality of quantum systems, highlighting the correlation between objective metrics and developers’ practical perceptions.

In conclusion, this research provides empirical support for the quantum component incorporated into the analyzability model. The goal is to show that indicators specific to quantum mechanics effectively represent circuit complexity and align with users' perceived complexities. Validation of these apsects is crucial for forming a solid and practical reference framework for actual quantum software, which will also facilitate integrating classical and quantum quality metrics into a unified model in the future.

\subsection{Research Questions (RQs)}

To empirically validate the analyzability model focused on quantum software, the following research questions have been formulated:

\begin{enumerate}
    \item \textbf{RQ1:} Is there a significant relationship between the analyzability levels obtained through the proposed model and quantum software developers' perception of quantum software complexity?
    \item \textbf{RQ2:} How do analyzability results vary when evaluating quantum systems in different experimental contexts and with diverse subject profiles?
    \item \textbf{RQ3:} To what extent does the quantum computing-specific component of the analyzability model enable a comprehensive evaluation of analyzability in quantum circuits?
\end{enumerate}

These questions guide the experimental design and result analysis, allowing for a precise assessment of the model's effectiveness and applicability in quantum software. The formulated research questions focus on integrating quantum-specific approaches in evaluating software quality. In this regard, studies, such as \cite{cerezo2021variational}, underscore the importance of using metrics tailored to quantum computing for accurately assessing quantum circuits. These findings reinforce the relevance of our approach by suggesting that incorporating quantum indicators can effectively complement traditional methods in measuring analyzability in quantum systems.

\subsection{Contributions}

This paper presents the following main contributions:

\begin{enumerate}
    \item \textbf{Empirical validation of the model:} A family of experiments has been designed and conducted to relate the analyzability levels calculated through the proposed model with the complexity perception reported by participants. The validation encompasses various experimental contexts and subject profiles, providing evidence of the correlation between objective metrics and subjective evaluations. This finding aligns with recent studies \cite{cerezo2021variational} and reinforces the model's applicability in real-world scenarios.

    \item \textbf{Identification of challenges and recommendations:} Based on the experimental analysis, the main methodological and technical challenges in applying quality metrics in quantum environments have been identified. Practical recommendations are provided to optimize the development and maintenance of systems integrating classical and quantum components, laying a solid foundation for future research and advancements in quantum software engineering.
\end{enumerate}

\subsection{Target Users and Practical Applications}

Quantum software offers many opportunities for different user profiles and practical applications in real-world scenarios. Two key dimensions are identified: target users and use cases.

The proposed model is designed for:
\begin{itemize}
    \item \textbf{Researchers and Academics:} Professionals and students in quantum computing and software engineering who seek to explore new methodologies for integrating classical and quantum paradigms, optimizing the quality and performance of complex systems \cite{Rönkkö2024}.
    \item \textbf{Quantum Software Developers and Engineers:} Experts in implementing solutions using quantum algorithms who can benefit from an evaluation framework that identifies critical areas for improvement in software design and maintenance \cite{Sepúlveda2025}.
    \item \textbf{Quality Assurance Specialists:} Professionals responsible for validating the design, maintainability, and quality of quantum systems, who will find in the model a tool to detect improvements and optimize diagnostic processes \cite{Pérez-Castillo20222375}.
    \item \textbf{Professionals in Various Sectors:} Innovation teams in industries such as finance \cite{Kea2024}, chemistry, healthcare \cite{Zhu2025}, etc., where the adoption of quantum technologies offers competitive advantages in terms of efficiency and processing capacity.
\end{itemize}

Additionally, the proposed quality model may have the potential to positively impact various domains:
\begin{itemize}
    \item \textbf{Optimization and Simulation:} In areas such as combinatorial optimization and the simulation of physical or chemical systems, the synergy between quantum and classical processing can reduce computation times and improve accuracy in solving highly complex problems \cite{Bilal2025}.
    \item \textbf{Cryptography and Security:} Integrating quantum techniques in developing cryptographic protocols can enhance communication security and provide powerful mechanisms against cyber-attacks \cite{Aardal2025497}.
    \item \textbf{Artificial Intelligence and Data Analysis:} The application of quantum models in machine learning enhances the processing of large volumes of data, facilitating pattern extraction and real-time decision-making \cite{Dutta2025}.
    \item \textbf{Development of New Products and Services:} In emerging industries, assessing the quality of quantum systems enables the rapid adoption and scalability of disruptive technologies, driving innovation in competitive markets \cite{Ouyang2025203}.
\end{itemize}

The proposed approach is designed to meet the needs of a broad range of users, from academia to industry. It offers practical applications that span optimization and simulation to security and data analysis improvements. This model bridges the gap between cutting-edge research and practical implementation, facilitating the transition toward increasingly integrated and efficient software systems.

\subsection{Novelty}

The main novelty of this work lies in the empirical validation of the quantum component of the proposed quality model through a family of experiments that integrate different participant profiles and experimental conditions. While the literature includes models for evaluating the quality of classical software \cite{Argotti2024}\cite{Gordieiev2024118875}\cite{Storey2025} and some preliminary studies addressing the evaluation of quantum systems \cite{Faryal2022}, this study is pioneering in combining both perspectives into a single evaluation framework. It incorporates traditional metrics (based on the ISO/IEC 25010 standard) alongside newly developed metrics designed explicitly for quantum circuits.

Our work focuses on the design and execution of a family of experiments that integrate diverse participant profiles and experimental conditions to empirically validate the quantum component of the proposed analyzability model. This experimental approach enables an objective assessment of how the structural clarity of circuits influences participant performance across different experiments. By considering contexts ranging from academic environments to researchers and industry professionals, our study bridges the gap between theory and practice, demonstrating that the model is practical under controlled conditions and applicable to real-world scenarios. This empirical validation lays the groundwork for using the model as a diagnostic and optimization tool in quantum software engineering, expanding the possibilities for continuous improvement in the quality and maintainability of these systems.

\subsection{Structure}

The remaining part of the research paper is organized as follows: Section~\ref{related_work} reviews related work and establishes the theoretical context; Section~\ref{model} presents the analyzability model for hybrid software, including the fundamentals of ISO/IEC 25010 and the integration of classical and quantum metrics; Section~\ref{sec: family design} describes the design of the experiment family, detailing the key differences between the validation experiments, as well as the variables, factors involved, and formulated hypotheses; Section~\ref{methodology} explains the methodology used for conducting the experiments and collecting data; Section~\ref{results} presents the descriptive statistical analysis and hypothesis testing, along with a comprehensive summary of the results; Section~\ref{discussion} addresses the research questions based on the obtained results and evaluates their limitations and challenges; Section~\ref{conclusions} provides the study's conclusions and outlines future research directions; and finally, Section~\ref{availability} details the availability of data and experimental materials.

\section{Related Work}
\label{related_work}

Research on evaluating quantum software quality and hybrid systems has experienced significant growth in recent years, with several studies addressing fundamental aspects of adapting traditional models to the specific characteristics of quantum computing.

The work of Gaur, Singh, and Ghanekar \cite{Gaur2019356} focuses on simplifying and modifying multi-controlled Toffoli circuits to improve their testability. This study highlights the importance of optimizing the design of complex circuits to facilitate their analysis and maintenance, which is essential in a context where quantum nature introduces additional challenges in terms of complexity and error control.

Subsequently, Cruz-Lemus, Marcelo, and Piattini \cite{Cruz-Lemus2021239} proposed a set of metrics to evaluate the \textit{understandability} of quantum circuits. Their research addresses the need to define indicators that objectively assess how easily a circuit can be understood, a crucial aspect for debugging and maintaining hybrid systems. This approach underscores the importance of adapting traditional quality metrics to the unique properties of quantum circuits.

In 2023, Kumar \cite{kumar2023quantumcomplexity} advanced the formalization of structural criteria for test case coverage in quantum software testing. His work integrates theoretical foundations and practical applications, offering a methodological framework that contributes to the validation of circuits in high-complexity environments and lays the groundwork for adopting specialized testing methods in quantum computing.

During 2024, Stefano, Di Nucci, Palomba, and De Lucia \cite{Stefano2024} conducted an empirical study focused on analyzing the effects of transpilation (transformation + compilation) on the emergence of “circuit smells” in quantum circuits. The results of this study demonstrate how compilation optimization processes can impact the quality of quantum design, affecting the maintainability and understandability of circuits. This raises the need to adjust these processes to improve the quality of hybrid software.

Furthermore, Kundu, Acharya, and Sarkar \cite{Kundu2024} proposed a scalable quantum gate-based implementation for conducting causal hypothesis testing. This work focuses on the practical application of quantum techniques in optimizing testing processes, demonstrating that hybrid approaches can effectively address complex problems in quantum computing.

Finally, Jang, Oh, Kim, and Seo \cite{Jang2024} developed a quantum implementation of a method aimed at achieving low-depth circuits, emphasizing the importance of reducing complexity without compromising efficiency. This study highlights the relevance of designing optimized circuits capable of operating at high speeds while efficiently utilizing quantum resources, which are critical for practical implementation in industrial environments.

Collectively, these studies have established the theoretical and practical foundations that justify integrating classical and quantum metrics into a unified quality evaluation framework for hybrid systems. The contributions of each study have advanced the understanding of how quantum circuit complexity and analyzability influence maintainability and performance, which is fundamental to developing the model validated in this article.

However, these studies highlight the need for an evaluation framework that integrates both classical and quantum metrics to comprehensively address the quality of hybrid software. While various metrics and optimization methods for quantum circuits have been proposed \cite{Cruz-Lemus2021239}\cite{kumar2023quantumcomplexity}, there is still a gap in the literature regarding the overall evaluation of analyzability in systems that combine both paradigms \cite{diaz2024environment}\cite{liu2022benchmarking}. This work is positioned within this context by proposing and empirically validating the quantum component of a hybrid quality model, offering an innovative approach to evaluating analyzability from an integrated perspective and laying the groundwork for future research and practical applications.

\section{The Analyzability Model for Hybrid Software}
\label{model}

\subsection{Fundamentals of ISO/IEC 25010 and Analyzability}

The ISO/IEC 25010 standard \cite{iso25010}, part of the ISO/IEC 25000 (SQuaRE) series \cite{iso25000}, provides a framework for software quality evaluation, defining eight essential characteristics: functionality, reliability, usability, efficiency, maintainability, portability, compatibility, and security. Each of these characteristics is further broken down into sub-characteristics that enable a more detailed assessment of software quality.

Within the scope of maintainability, \textit{analyzability} is a key sub-characteristic, as it determines how easily software can be understood and diagnosed. Evaluating analyzability involves measuring code clarity, the ease of identifying errors, and the ability to make modifications with a predictable impact. In classical software, analyzability is measured using metrics such as cyclomatic complexity, module structuring, and code documentation \cite{rodriguez2014software}. The importance of this sub-characteristic lies in the fact that software with low analyzability tends to be more challenging to maintain and evolve, increasing development costs and raising the likelihood of errors.

However, the direct application of these approaches to hybrid software is limited due to the fundamental differences between classical and quantum computing.

\subsection{Analyzability of Classical Software}
\label{subsec: class software analyzability}

To evaluate analyzability in classical software, various metrics and principles are employed to measure code clarity, architectural modularity, and ease of maintenance \cite{rodriguez2014software}. Below are the main metrics used in classical environments to assess this sub-characteristic:

\begin{enumerate}
    \item \textbf{Coding Rules:} These are established guidelines for software development that ensure code readability and comprehensibility. These rules include naming conventions, code structures, and design principles. Well-structured code that adheres to these rules facilitates understanding and reduces errors.

    \item \textbf{Code Documentation:} Documentation is essential for clearly explaining the software’s functionality, purpose, and behavior. \textit{In-line} comments, external documents such as \texttt{README} files, and design specifications play a fundamental role in analyzability. Well-documented code is more manageable for other developers to understand and modify.

    \item \textbf{Cyclomatic Complexity:} Introduced by McCabe \cite{mcabe1976complexity}, cyclomatic complexity measures the complexity of a program's control flow, precisely the number of independent paths through the code. Higher complexity makes the code more difficult to understand and modify. Reducing cyclomatic complexity improves analyzability by facilitating a clearer understanding of software behavior.

    \item \textbf{Package Structuring:} The organization of modules and packages within the software directly impacts its analyzability. A well-structured system allows developers to quickly understand relevant modules without comprehending the entire system. The hierarchy and modularity of packages are also essential for maintaining clarity.

    \item \textbf{Class Structuring:} In object-oriented programming, the organization of classes significantly influences analyzability. Well-designed classes with clear responsibilities are easier to understand. The \textit{Single Responsibility Principle} (SRP) and low coupling between classes are key characteristics of good structuring.

    \item \textbf{Method Size:} Overly long methods or functions are often more challenging to understand. Keeping methods small and focused on a single task improves code analyzability. It is recommended that a method should not exceed 10-15 lines of code to facilitate comprehension and modification.

    \item \textbf{Duplicated Code:} Code duplication is one of the primary sources of difficulty in maintainability and analyzability. Duplicated code increases system complexity and makes modifications more challenging, as changes must be replicated across multiple parts of the code. Eliminating duplicated code through techniques such as refactoring significantly improves analyzability.
\end{enumerate}

The application of these metrics has proven to be effective in assessing software quality in industrial environments \cite{Verdugo2024Laboratory}. Various studies have confirmed that improving code analyzability reduces the effort required for software maintenance and decreases the likelihood of errors and failures in production.

It is worth noting that, although the empirical validation presented in this manuscript focuses exclusively on quantum circuits, the overall analyzability model is hybrid in nature and was designed to be applicable to systems that combine both classical and quantum components. Consequently, the classical software metrics presented were not directly employed in the evaluation described in this study, as the objects under analysis were purely quantum in structure.

However, these classical metrics constitute a core part of the proposed hybrid model and have already been validated in traditional software contexts. In particular, they are part of the metric set routinely applied by the first and, to date, only ISO/IEC 25000-accredited software quality laboratory, of which the authors are members \cite{Verdugo2024Laboratory}. Therefore, the absence of classical metrics in the present empirical validation is due to the quantum-only nature of the analyzed artifacts, and not to a lack of prior validation of those metrics.

In future research, we plan to extend the empirical validation to hybrid artifacts that integrate classical and quantum modules (e.g., orchestrated via microservices), where both sets of metrics will be applied in parallel. This will enable a comprehensive interpretation of analyzability and the study of interactions between classical and quantum complexity patterns within the same system.

\subsection{Analyzability of Quantum Software}
\label{subsec: quantum software analyzability}

Quantum computing, in turn, is based on development paradigms that differ fundamentally from those of classical systems and cannot be directly evaluated using traditional metrics. Since quantum circuits operate with qubits in superposition and entanglement, and their execution is probabilistic, their analyzability cannot be measured solely with conventional approaches. This creates the need to develop new quantum-specific metrics while adapting those applicable in quantum and hybrid environments.

The metrics proposed in the analyzability model for hybrid software, which we validate in this article, are as follows:

\begin{enumerate}
    \item \textbf{Circuit width}: Measures the number of qubits used in a quantum circuit at any given time. A circuit with greater width may be more challenging to understand and debug as it involves more qubits, increasing its complexity.

    \item \textbf{Circuit depth}: Represents the number of layers of quantum gates applied in a circuit before obtaining the final result. Greater depth implies increased execution complexity, which can affect analyzability and raise the likelihood of errors due to qubit interference.

    \item \textbf{Gate complexity}: Evaluates the number and type of quantum gates used in the circuit. Some gates, such as Hadamard or phase gates, are more straightforward, while others, such as controlled or swap gates, increase circuit complexity and analysis difficulty.

    \item \textbf{Conditional instructions}: Indicates the number of quantum operations whose behavior depends on the state of other qubits. Conditional instructions can make the circuit harder to understand by introducing dependencies and non-trivial behaviors during execution.

    \item \textbf{Quantum cyclomatic complexity}: Adapting classical cyclomatic complexity to quantum circuits. It is based on the number of independent paths that can be taken during circuit execution. A high value suggests greater difficulty in analyzing and debugging the circuit \cite{kumar2023quantumcomplexity}.

    \item \textbf{Measurement operations}: Measures how often a measurement is performed in a quantum circuit. Many measurements can make the circuit harder to understand, primarily if they depend on complex conditions.

    \item \textbf{Initialization and reset operations}: Represents the number of times qubits are prepared in a specific state before or after other quantum operations. Excessive use of these operations can increase the overall complexity of the circuit.

    \item \textbf{Auxiliary qubits}: Measures the number of additional qubits required to execute a circuit. A higher number of auxiliary qubits may indicate greater structural complexity, making the circuit harder to analyze.
\end{enumerate}

Additionally, several studies \cite{AlvaradoValiente2023QuantumServices}\cite{diaz2024QQservices} have applied some of these metrics in controlled environments to evaluate the quality of quantum algorithms. In particular, metrics such as quantum cyclomatic complexity, circuit depth, and gate complexity have proven to be practical tools for analyzing the difficulty of understanding and debugging quantum circuits. These studies have demonstrated that quantum metrics can objectively evaluate the structural behavior of circuits, enabling the identification of high-complexity patterns that could impact the efficiency and reliability of algorithms. Using these metrics has also facilitated the early detection of issues in circuit design, contributing to improved maintainability of quantum software and optimizing the development of quantum and hybrid applications.

\section{Designing the Family of Experiments}
\label{sec: family design}

Although the proposed model is designed to support the quality evaluation of hybrid software systems—that is, systems combining classical and quantum components—this validation focuses exclusively on the \textit{quantum software} component. More specifically, it targets \textit{quantum circuits}, which represent a well-established paradigm within quantum computing and are widely used in both theoretical and applied contexts to describe quantum algorithms at the level of qubits, gates, and measurements.

Among the various paradigms in quantum computing—such as quantum Turing machines, measurement-based models, and quantum annealing—the circuit-based approach is currently the most widespread and supported by leading frameworks like Qiskit (IBM) and Quirk. This paradigm models computation as a sequence of quantum gates acting on qubits and is particularly well suited for structural analysis. Consequently, the analyzability model presented in this work was specifically designed for gate level quantum circuits.

The programs used in this study are representative of this paradigm: circuit like quantum programs written in a low-level, assembly-like syntax analogous to OpenQASM, where gates and arguments are hard coded and the qubit layout is explicitly defined. This design choice ensures that core model metrics such as depth, width, gate complexity, and qubit interactions can be extracted and interpreted in a consistent and reproducible manner.

We acknowledge that application-oriented and multi-subroutine quantum programs are gaining attention due to their modularity and closer alignment with classical programming practices. These high-level constructs, explored in recent studies such as \cite{long2024}\cite{lubinski2023} , will be considered in future extensions of our model. However, for this initial empirical validation, circuit-based quantum software provides a controlled and interpretable foundation for evaluating the relationship between structural complexity and understandability.

To validate the quantum component of the proposed analyzability model for hybrid software, we conducted a family of experiments involving participants with diverse backgrounds in quantum computing and software quality. This approach enables the collection of more representative results by evaluating the model across different educational and professional contexts.

The experiments were designed to assess the model’s effectiveness in measuring how easily quantum circuits can be understood and analyzed. Specifically, we investigated the relationship between the analyzability level of a series of circuits and the scores obtained by participants in exercises measuring their comprehension. By incorporating four distinct validation experiments, we were able to compare the model’s performance across groups with varying levels of experience, thereby strengthening its external validity.

\begin{figure}[htbp]
  \centering
  \includegraphics[width=1\textwidth]{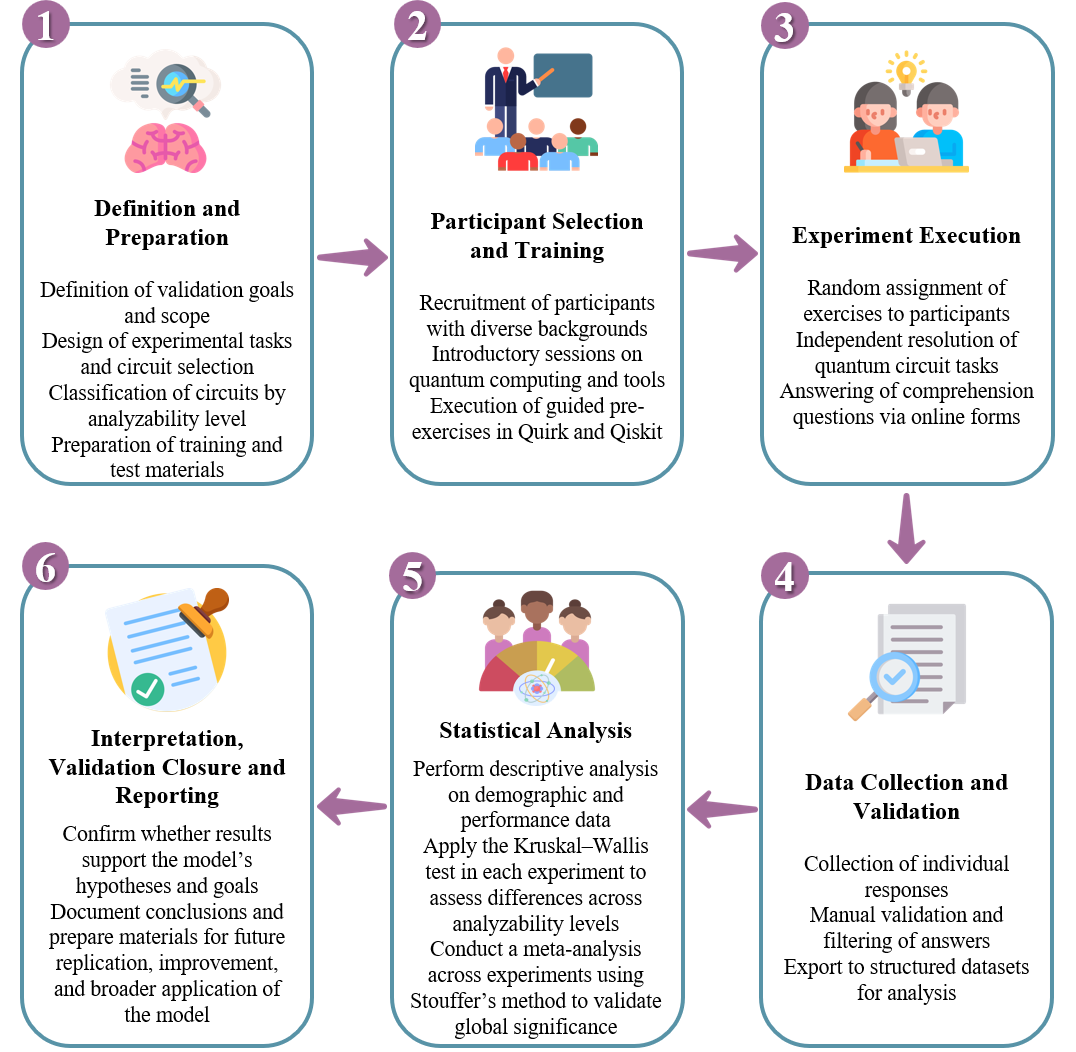}
  \caption{Validation workflow summarizing the phases of the experimental process}
  \label{fig:workflow_validation}
\end{figure}

To provide a comprehensive overview of the validation approach adopted in this work, Figure~\ref{fig:workflow_validation} summarizes the complete workflow followed throughout the experimental process. This diagram outlines all key phases, from the initial design and preparation of materials to participant training, execution, data analysis, and interpretation of results. The goal is to offer a clear visualization of the methodological rigor applied, emphasizing how each stage contributes to validating the quantum component of the analyzability model across diverse user profiles.

\subsection{Experiment Features and Characteristics}

The four experiments were designed to evaluate the model in different contexts, using participant groups with diverse characteristics in terms of academic background and experience levels in classical and quantum computing. Below, the main distinguishing aspects of each experiment are described:

\begin{itemize}
    \item \textbf{1-MSc: Master students in Computer Science (University of Bari - Aldo Moro)}  
    This experiment involved 14 subjects enrolled in master's programs at the University of Bari Aldo Moro. This group represents an intermediate level of experience in quantum computing and software quality, as, although they are not experts in the field, they have an advanced background in computer engineering. Evaluating this group allows us to observe how the model performs in an academic setting with participants with prior knowledge but no specialization in quantum software quality.

    \item \textbf{2-BScMay: Undergraduate students in Computer Science (University of Bari - Aldo Moro)}  
    This experiment involved 84 subjects enrolled in the third year of a computer science undergraduate program at the University of Bari Aldo Moro. Unlike the first experiment, these students have a more basic educational background and minimal experience in quantum computing and software quality. Validating the model with this group allows us to analyze how it performs when applied to subjects without prior knowledge, providing key insights into its ability to differentiate circuits with varying levels of analyzability without bias from advanced technical expertise.

    \item \textbf{3-Prof: Subjects experienced in quantum computing and software quality (University of Castilla-La Mancha, University of Extremadura, and University of Deusto)}  
    This experiment included 15 participants from the University of Castilla-La Mancha (UCLM), the University of Extremadura (UEx), and the University of Deusto (UD), all of which are Spanish universities. Unlike the other experiments, this study was conducted online. The group comprised a heterogeneous mix of profiles, including junior and senior researchers, technologists and doctoral fellows, and industry professionals, all of whom had some previous experience with quantum computing or software quality. Notably, two of the participants—one male and one female—were in the 18–21 age range. Both had been actively involved in research projects related to quantum software from an early stage, which justified their inclusion in this group.

    \item \textbf{4-BScNov: Undergraduate students in Computer Science (University of Bari - Aldo Moro)}  
    This experiment involved 109 subjects from the University of Bari Aldo Moro enrolled in a computer science undergraduate program. This group is similar to the 2-BScMay experiment but with a larger sample, increasing the representativeness of the results for subjects with a university-level of education but without specialization in quantum computing. Comparing the results of this group with those obtained in the 2-BScMay experiment allows for an analysis of the model’s consistency when applied to a similar population at different points in the study.
\end{itemize}

The differences in education and experience among these groups allow for the evaluation of the model in diverse scenarios, leading to a more robust validation and strengthening the study’s validity. Table~\ref{tab:experiments_info} summarizes the key characteristics of the experiments.

\begin{table}[htb]
\centering
\caption{Key information of the experiments}
\label{tab:experiments_info}
\begin{tabular}{lccccc}
\toprule
\textbf{Experiment} & \textbf{N} & \textbf{Profile} & \textbf{Mode} & \textbf{Date} \\
\midrule
1-MSc     & 14  & Master's Students      & On-site. Bari (Italy) & May 2024 \\
2-BScMay  & 84  & Undergraduate Students & On-site. Bari (Italy) & May 2024 \\
3-Prof    & 15  & Subjects experienced in in QC and QS & On-line & June 2024 \\
4-BScNov  & 109 & Undergraduate Students & On-site. Bari (Italy) & Nov. 2024 \\
\bottomrule
\end{tabular}
\end{table}

\subsection{Subjects Demographics}

\begin{figure}[htbp]
  \centering
  \includegraphics[width=1\textwidth]{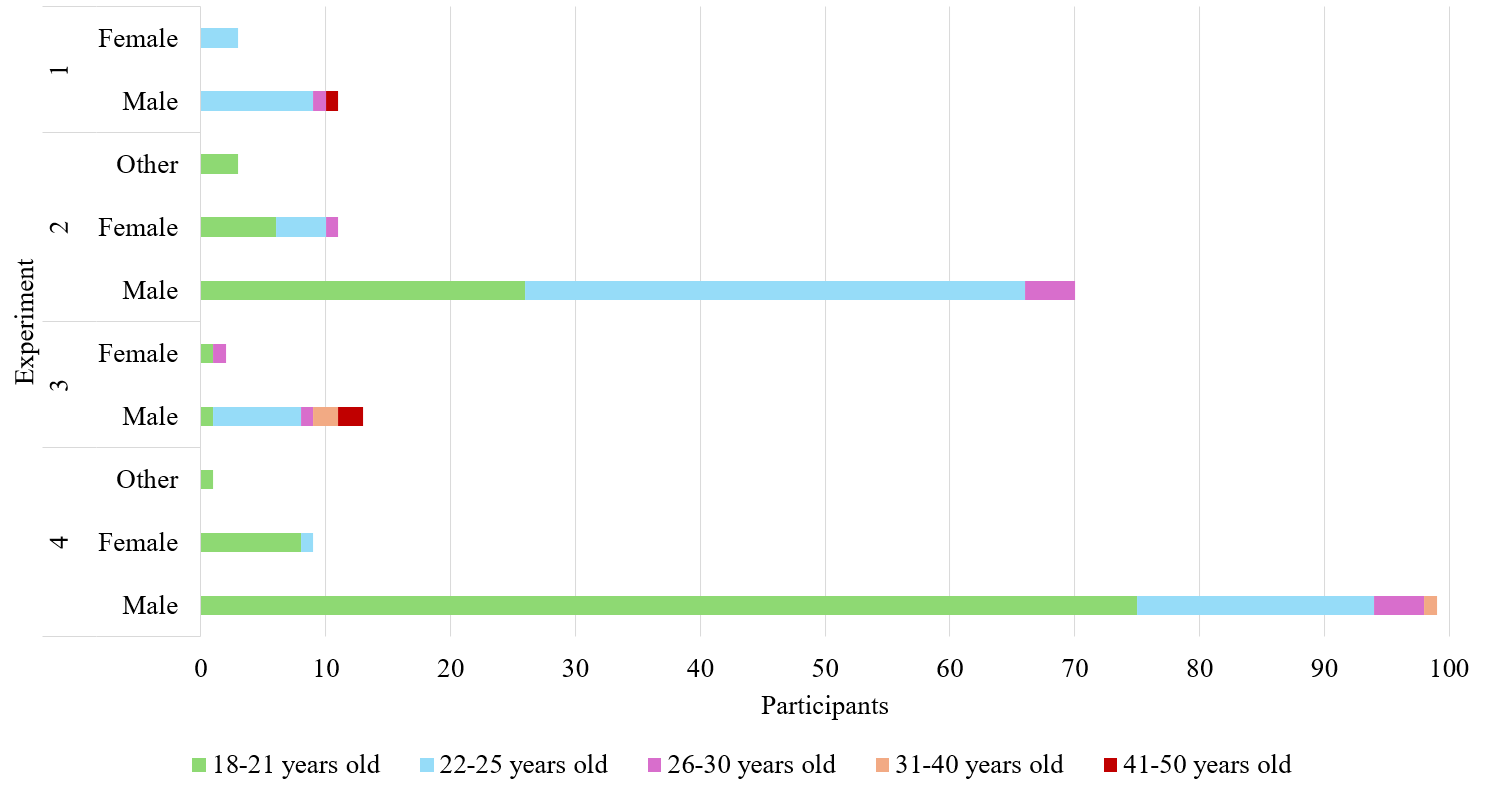}
  \caption{Distribution of all experiments by gender and age groups}
  \label{fig:distrib}
\end{figure}

The distribution of participants by gender and age in each experiment is detailed in Figure~\ref{fig:distrib}.  

In \textbf{1-MSc}, most subjects fall within the 21 to 24 age group, with a distribution of 3 women and 9 men, followed by a small group of participants aged 24 to 50, consisting exclusively of men. In \textbf{2-BScMay}, the largest group belongs to the 18 to 21 age range, with a male majority (26 men, 6 women, and 3 individuals of another gender). The trend continues in the 22 to 25 age group, with an apparent disparity of 40 men and 4 women. This experiment also recorded some participants in the 26 to 30 age range, although in a smaller proportion. The trend remains in the subsequent age groups' participation rate, with a more balanced distribution across age groups. A male predominance is observed in the 22 to 25 age range, while in the older age groups (26 to 50), all participants were men except for one woman in the 26 to 30 group. In \textbf{4-BScNov}, the 18 to 21 age range was once again the most represented, with 75 men, 8 women, and 1 individual of another gender. In the subsequent age groups, the trend remains, with a male majority in each age category.

Analyzing all experiments together (see Figure~\ref{fig:distrib}), it is confirmed that the majority of the sample consists of young participants aged between 18 and 21, with a clear male predominance across all age groups (102 participants), compared to women (15 participants) and individuals who identified with another gender (4 participants). In the following age range (22 to 25 years), the trend remains, with a higher proportion of men (66) compared to women (5). From age 26 onward, the number of participants decreases significantly, with only a few records in each category.

This gender disparity may be attributed to various factors, such as the historically lower representation of female role models or differences in vocational orientation towards technological disciplines. Despite numerous studies and initiatives aimed at understanding and addressing these issues, the data reflect that the gender gap in Computer Engineering remains significant \cite{Cheryan2017}. These results underscore the need to continue implementing effective strategies that promote greater participation of women and gender-diverse individuals in this field.

In line with the current European guidelines for good scientific practice and gender equality, such as those established in Horizon Europe \cite{HorizonEurope} and the Gendered Innovations framework \cite{GenderedInnovations}, we have chosen to dis-aggregate the results by gender across all experimental groups, even in cases where the number of female or gender-diverse participants is small. While such small samples do not allow generalizable conclusions, their inclusion aligns with principles of transparency and inclusion in scientific reporting, ensuring that all represented groups are acknowledged in the analysis.

The influence of external factors has been controlled through the random assignment of subjects to different circuits, minimizing biases in the evaluation. However, including multiple groups with diverse characteristics allows a comparative analysis between subjects with different levels of education and experience, providing greater capability for generalizing the results obtained.

\subsection{Description of Experimental Materials}
\label{sub-sec:materials}

To conduct the experimental tasks, various circuits were used in each of the experiments. Each circuit was assigned an analyzability level by applying a hybrid software quality model proposed in several previous works, using a metric-based evaluation environment specifically designed for quantum and classical systems \cite{diaz2025environment}\cite{diaz2024environment}\cite{diaz2024metrics}. This automated environment analyzes each circuit’s structural properties—such as depth, width, gate complexity, and quantum cyclomatic complexity—and calculates an analyzability score according to the model. This model is part of a whole evaluation framework which builds upon over two decades of experience in software quality evaluation and certification, as detailed in \cite{Verdugo2024Laboratory}.

Although the model includes additional metrics—such as auxiliary qubits, reset operations, or conditional instructions—these were not explicitly activated for computing the analyzability levels of the circuits used in this validation. This decision was made to ensure interpretability and reduce complexity in the experimental setting. In particular, auxiliary qubits were deliberately avoided, as their purpose and behavior are more difficult to explain to participants without advanced quantum programming knowledge, and they are rarely present in canonical quantum algorithms.

Despite this simplification, all circuits employed in the experiments were formally evaluated using the model, and their final analyzability level (1/5, 3/5, or 5/5) was assigned based on the metric-based score computed by the automated environment. The evaluation criteria—including thresholds, weightings, and level definitions—are detailed in previous publications that formally introduce and validate the model \cite{diaz2025environment}\cite{diaz2024environment}\cite{diaz2024metrics}, and were therefore not repeated in full here.

It should be noted that the exact configuration and scoring functions used by the model are not publicly disclosed, as the model is intended for future commercialization to support quality certification of hybrid quantum–classical software systems. However, the conceptual foundations, core metrics, and evaluation strategy are transparently documented in the cited works, and are consistent with quality assurance methodologies that have been successfully applied to classical systems in industrial settings.

\begin{itemize}
    \item \textbf{Low analyzability:} The circuit was characterized by high structural complexity; it included many complex quantum gates, significant depth, and width and exhibited a high level of cyclomatic complexity. According to the model, these circuits have an analyzability score of 1/5.
    \item \textbf{Medium analyzability:} This circuit balanced complexity and clarity, allowing for a differentiated evaluation of the effect of structural variability on the perception of analyzability. These circuits had a score of 3/5 in the model.
    \item \textbf{High analyzability:} The circuit was designed to maximize clarity and structural simplicity. It featured fewer operations, lower depth and width, and an optimized arrangement of qubits. According to the model, these circuits have an analyzability level of 5/5.
\end{itemize}

Originally, the experimental design contemplated five levels of analyzability as described in the Registered Report \cite{ana2024validationanalysabilitymodelhybrid}. However, for the final implementation, the number of levels was reduced to three. This decision was made to reduce the cognitive load and fatigue faced by the participants, due to the inherent difficulty of the tasks to be performed.

Each circuit was accompanied by a set of eight questions designed to evaluate participants’ ability to read, interpret, and reason about quantum circuits. These questions included both multiple-choice and open-ended formats, covering aspects such as gate identification, circuit structure, output states, probability distributions, and specific quantum properties (e.g., measurement outcomes or amplitudes for given states). The complete experimental material and all data related to the analysis presented later are available online in a public repository\footnote{\url{https://github.com/aniitadiazm/EmSE2025-Diazetal}}.

\subsection{Variables and Factors}

The experimental design considers various variables and factors that may affect the results obtained in the four validation experiments. These variables allow for evaluating the effectiveness of the analyzability model in measuring quantum circuits with different levels of complexity and determining the extent to which external factors may influence participants' interpretation and understanding of the circuits.

\textbf{Independent Variable}:  
The independent variable in the experiment is the \textbf{analyzability level of the quantum circuit}, as determined by the proposed model. This variable takes on the three values described in subsection \ref{sub-sec:materials}: low, medium, and high analyzability.

\textbf{Dependent Variable}:  
The dependent variable in the experiment is the \textbf{score obtained by participants in the evaluation of the circuits}. This score reflects each participant’s performance in reading, understanding, and analyzing quantum circuits with different levels of analyzability.

For each circuit, participants were asked to answer a total of eight questions that assessed various comprehension aspects, including:
\begin{itemize}
    \item identification of output states
    \item calculation of amplitude probabilities provided by the circuit simulator via the \texttt{mag$^2$} parameter values
    \item recognition of applied quantum operations
    \item interpretation of the circuit’s overall functionality
\end{itemize}


These scores are normalized to a percentage scale ranging from 0\% (no correct answers) to 100\% (all correct answers), enabling direct comparison across different circuits and participant groups. For example, a raw score of \(6/8\) would be normalized as 75\%.

In addition to the relationship between circuit analyzability and the scores obtained by participants, several factors have been identified that could influence the experimental results. These factors include:

\begin{itemize}  
    \item \textbf{Previous experience in quantum software:} Participants with more significant experience in quantum computing are expected to have a higher capacity to analyze more complex circuits and achieve better exercise scores.  
    \item \textbf{Previous experience in classical software:} Familiarity with traditional systems may influence the speed and accuracy with which subjects comprehend logical structures and analyze circuits.  
    \item \textbf{Educational level:} Subjects with a higher educational level may exhibit greater abstraction and analytical skills, potentially facilitating their understanding of quantum circuits.  
    \item \textbf{Order of circuit analysis:} The sequence in which participants analyze quantum circuits may affect their performance. For instance, if a subject starts with circuits of low analyzability before encountering more complex ones, they might develop an analysis strategy that helps improve their performance. Conversely, if they begin with highly complex circuits, the difficulty might affect their initial performance, influencing their perception of subsequent circuits.
\end{itemize}  

Incorporating these factors into the experimental design allows for a more detailed analysis and a better understanding of how they influence participants' ability to analyze quantum circuits with different levels of analyzability.

In future replications, additional variables such as \textbf{response time} and \textbf{self-perceived difficulty} could also be considered to triangulate participants’ performance and further validate the impact of analyzability levels.

\subsection{Hypotheses}

The primary objective of the experiment is to evaluate the relationship between the analyzability level of quantum circuits, as determined by the proposed model, and the participants' performance in terms of circuit comprehension and analysis. To this end, the following hypotheses are formulated:

\textbf{Null Hypothesis (\(H_0\))}:  
There is no statistically significant relationship between the analyzability level of quantum circuits and the average score obtained by participants. In other words, the analyzability level computed from the model does not determine the ease with which circuits can be understood and analyzed.

\textbf{Alternative Hypothesis (\(H_a\))}:  
There is a significant relationship between the analyzability level of quantum circuits and the average score obtained by participants.

The empirical validation through the family of experiments will determine whether the relationship between the model's analyzability metric and participant performance holds across different subject groups with varying levels of experience and educational backgrounds. The following section provides a detailed description of the participants' characteristics in each validation experiment.

\section{Performing the Family of Experiments}
\label{methodology}

Before conducting each experiment, participants attended a couple of training sessions to gain essential knowledge of quantum computing and quantum circuit design. It is important to note that while some participants received comprehensive theoretical training covering the fundamental principles of quantum computing—including the nature and behavior of qubits, primary operations using quantum gates, and the process of quantum state measurement—the group consisting of professionals, software experts, and subjects experienced in quantum computing did not receive such prior training; they were only given a brief introduction and explanation of what they would encounter. Additionally, practical examples of circuit construction and simulation were presented using specialized tools such as \texttt{IBM Qiskit}\footnote{\url{https://www.ibm.com/quantum/qiskit}} and \texttt{Quirk}\footnote{\url{https://algassert.com/quirk}}, which were later incorporated into the experimental phase.

To ensure that all participants correctly understood the necessary concepts and familiarized themselves with the experiment's dynamics, guided pre-exercises were conducted. These exercises, although less complex than those included in the experimental test, allowed participants to gain experience with the tools and methodologies they would later use, minimizing any potential biases resulting from a lack of prior knowledge of quantum computing or the use of simulators.

During the experimental phase, a structured data collection system was designed to ensure the randomization of tasks and the individualized tracking of participant responses. To achieve this, each participant received a personalized \texttt{PDF} document via email, which contained three access links to the exercises. These exercises were assigned randomly to prevent potential effects such as fatigue or progressive learning, which could alter the results.

The exercises were based on two main tools:

\begin{itemize}
    \item \textbf{Qiskit Code:} Participants were provided with \texttt{Qiskit} code snippets that they had to interpret to analyze the behavior of the corresponding quantum circuit.
    \item \textbf{Quirk Simulator:} Subjects were required to replicate the circuits in the \texttt{Quirk} simulator, an open-source, browser-based tool for designing and simulating quantum circuits. Quirk enables real-time visualization of the evolution of quantum states, including amplitudes, phase rotations, and entanglement. By observing the output states and corresponding probability amplitudes (mag$^2$ parameter), participants could analyze the functional behavior of each circuit. This facilitated a deeper understanding and served as the basis for answering the comprehension questions in the evaluation.
\end{itemize}

The primary objective of these tasks was to assess participants' ability to understand the structure of quantum circuits and accurately predict the final states of qubits after execution. Once the circuit was correctly implemented in \texttt{Quirk}, participants were required to answer eight questions specifically designed to evaluate their comprehension of the circuit’s behavior. These required identifying the number of qubits, recognizing the applied operations (e.g., Hadamard, CNOT, or phase gates), interpreting output states, and calculating the probability amplitudes (mag$^2$) for specific results shown in the simulator. Correctly answering these questions required a complete understanding of the circuit logic—from initialization to gate application and measurements—as well as the ability to analyze the simulation output generated by \texttt{Quirk}.

Although \texttt{Quirk} provides intuitive visual feedback, it was used exclusively to replicate the circuits and observe their results; it did not offer automated solutions to the questions. The design of the questions demanded analytical reasoning beyond simple visual interpretation. Furthermore, the guided training sessions conducted before the experiments ensured that participants developed a solid conceptual understanding of quantum operations, helping to minimize over-reliance on the simulator’s interface.

The responses were recorded using \texttt{Google Forms}, which provided a structured digital environment to ensure consistency across all participants.  This setup ensured that the collected scores reflected actual circuit comprehension, not rote memorization or superficial knowledge. All questionnaires used in the experiments are publicly available in the project repository.

Upon completion of the experimentation phase, a data validation process was conducted, in which all responses were manually reviewed to identify and remove any that were incomplete or inconsistent. Subsequently, the data was exported in \texttt{CSV} format and converted into \texttt{Excel} files to facilitate processing and statistical analysis. Methodological consistency across different experiments minimized the influence of external factors, ensuring that the obtained data was reliable and comparable under each of the established experimental conditions.

A descriptive statistical approach was used for data analysis, including measures of central tendency and dispersion. Additionally, the non-parametric \textit{Kruskal-Wallis} test was applied to compare differences between the experimental groups. A more detailed analysis of the results and the statistical tests performed is presented in Section~\ref{results}.

\section{Analyzing the Family of Experiments}
\label{results}

A comparative analysis was conducted across the four previously described validation experiments to evaluate the effectiveness of the analyzability model in different contexts. First, a descriptive statistical analysis was performed on the participants' scores based on demographic variables, computing experience, and the order in which they received the circuits. This analysis identified general trends and preliminary differences among the experiments, exploring how factors such as age, educational level, or experience in quantum computing may influence participant performance.

\sloppy
Subsequently, a hypothesis test was conducted using the Kruskal-Wallis test \cite{kruskal1952}\cite{ostertagova}\cite{sherwani2021analysis}, which demonstrated that, when considering all subjects and circuits globally, statistically significant differences exist among the three analyzability levels proposed by the model. All non-parametric statistical tests performed were two-sided, following standard practice for the Kruskal-Wallis test. This choice is consistent with the non-directional nature of our hypotheses, which aimed to explore whether differences existed among analyzability levels without assuming a specific direction of effect. To integrate these results, a weighted meta-analysis of the four experiments was performed, combining the \(p\)-values obtained in each Kruskal-Wallis test using the weighted Stouffer method ~\cite{kim2013stouffer}\cite{Stouffer1949}\cite{Yoon2021}.

\subsection{Descriptive Statistical Analysis}

\subsubsection{Descriptive statistical analysis of 1-MSc}

To examine the distribution of the scores obtained in 1-MSc, Table~\ref{tab:descriptive_experiment1} can be referenced. From left to right, the columns contain the following data: the analyzability level of the circuit—where Low corresponds to circuits with an analyzability level of 1/5, Medium to circuits with an analyzability level of 3/5, and High to circuits with an analyzability level of 5/5—, the number of subjects in 1-MSc, the mean score obtained by participants, and the standard deviation of the scores.

The 1-MSc experiment was conducted with fourteen participants enrolled in a master's program in computer science, all of whom had an advanced education and several years of experience in classical programming. However, almost none had prior experience in quantum computing. To examine the distribution of the scores obtained, Table~\ref{tab:descriptive_experiment1} shows that the values of the \textbf{independent variable}—circuit analyzability—were categorized as low, medium, and high (1, 3, and 5/5, respectively, according to the model). In contrast, the \textbf{dependent variable}, that is, the score obtained by the subjects, is presented as the mean value and standard deviation.

\begin{table}[htb]
\centering
\caption{Descriptive statistics for 1-MSc scores}
\label{tab:descriptive_experiment1}
\begin{tabular}{lcccc}
\toprule
\textbf{Quality level} & \textbf{N} & \textbf{$\overline{x}$} & \textbf{std. dev.} & \textbf{95\% CI} \\
\midrule
Low    & 14 & 93.75\%  & 13.98\% & [85.68\%, 100.00\%] \\
Medium & 14 & 92.86\%  & 13.95\% & [84.81\%, 100.00\%] \\
High   & 14 & 100.00\% & 0.00\%  & [100.00\%, 100.00\%] \\
\bottomrule
\end{tabular}
\end{table}

Table~\ref{tab:descriptive_experiment1} summarizes the descriptive statistics of the scores obtained in 1-MSc, where 14 master's students participated. The quantum circuits used in this experiment were categorized into three levels of analyzability: low, medium, and high, based on their structural complexity and clarity.

The results indicate that low-analyzability circuits yielded an average score of 93.75\% (SD = 13.98\%), while medium-analyzability circuits achieved a slightly lower average of 92.86\% (SD = 13.95\%). High-analyzability circuits consistently received a perfect score of 100.00\%, with no variability (SD = 0.00\%), indicating uniform success across all participants for that condition.

These findings suggest that the advanced academic background of the participants may have mitigated the expected differences in circuit comprehension across analyzability levels. Despite the model's prediction that greater structural clarity would facilitate understanding, the homogeneity in scores indicates that this cohort was capable of interpreting even structurally complex circuits with high accuracy.

It should be noted that, due to the relatively low number of subjects in this experiment, the randomization of treatments may not have worked optimally. This situation is similar to the 3-Prof experiment.

The results presented in Table~\ref{tab:descriptive_by_gender_1-MSc} show the descriptive statistics of the scores based on gender for the 1-MSc experiment. In the male group (N = 11), the mean score for low-analyzability circuits is 94.32\%, with a standard deviation of 15.18. In contrast, for medium-analyzability circuits, the mean is 90.91\% (SD = 15.91\%), and for high-analyzability circuits, a perfect score of 100.00 is achieved with no variability. On the other hand, in the female group (N = 3), the recorded mean scores are 91.67\% (SD = 14.43\%) for low-analyzability circuits, 100.00 for medium-analyzability circuits, and 100.00 for high-analyzability circuits, with no variation in the latter two cases.

\begin{table}[htb]
\centering
\caption{Descriptive statistics by gender for 1-MSc scores}
\label{tab:descriptive_by_gender_1-MSc}
\begin{tabular}{lcccccccc}
\toprule
\textbf{Gender} & \textbf{N} & \multicolumn{2}{c}{\textbf{Low}} & \multicolumn{2}{c}{\textbf{Medium}} & \multicolumn{2}{c}{\textbf{High}} \\
\cmidrule(lr){3-4} \cmidrule(lr){5-6} \cmidrule(lr){7-8}
 &  & $\overline{x}$ & std. dev. & $\overline{x}$ & std. dev. &$\overline{x}$ & std. dev. \\
\midrule
\textbf{Male} & 11  & 94.32\% & 15.18\% & 90.91\% & 15.91\% & 100.00\% & 0.00\% \\
\textbf{Female} & 3  & 91.67\% & 14.43\% & 100.00\% & 0.00\% & 100.00\% & 0.00\% \\
\bottomrule
\end{tabular}
\end{table}

These findings indicate that, in general, both groups achieve very high scores in high-analyzability circuits, suggesting that structural clarity in these circuits facilitates comprehension uniformly. However, in the case of low- and medium-analyzability circuits, a slight difference in the mean scores is observed: men exhibit slightly higher scores compared to women in the low-analyzability category, while in the medium-analyzability category, the female group attains the maximum score. Nonetheless, it is essential to note that the female sample size is very small. This difference, although moderate, could be influenced by individual variability and the limited number of female participants, preventing definitive conclusions from being drawn. Overall, the general trend supports the validity of the analyzability model, demonstrating that clarity in quantum circuit design translates into improved performance regardless of gender. However, future research with more balanced samples could provide a more comprehensive perspective on these differences.

\subsubsection{Descriptive statistical analysis of 2-BScMay}

2-BScMay, primarily composed of undergraduate students, provides a complementary perspective by analyzing how the analyzability model performs in a group with less advanced training than 1-MSc. Table~\ref{tab:descriptive_experiment2} presents the statistical values describing the scores obtained for the three levels of analyzability (low, medium, and high).

\begin{table}[htb]
\centering
\caption{Descriptive statistics for 2-BScMay scores}
\label{tab:descriptive_experiment2}
\begin{tabular}{lcccc}
\toprule
\textbf{Quality level} & \textbf{N} & \textbf{$\overline{x}$} & \textbf{std. dev.} & \textbf{95\% CI} \\
\midrule
Low    & 84 & 89.29\% & 15.20\% & [86.04\%, 92.54\%] \\
Medium & 84 & 91.07\% & 12.73\% & [88.21\%, 93.92\%] \\
High   & 84 & 98.81\% & 10.85\% & [96.50\%, 100.00\%] \\
\bottomrule
\end{tabular}
\end{table}

Table~\ref{tab:descriptive_experiment2} summarizes the descriptive statistics of the scores obtained in 2-BScMay, where 84 undergraduate students participated. The quantum circuits used in this experiment were classified into three analyzability levels: low, medium, and high, based on their structural complexity and clarity.

According to the data, circuits with low analyzability yielded an average score of 89.29\% (SD = 15.20\%), while medium-analyzability circuits resulted in a slightly higher average of 91.07\% (SD = 12.73\%). High-analyzability circuits reached an even higher mean score of 98.81\% with a standard deviation of 10.85. These results support the hypothesis that increased structural clarity facilitates comprehension and analysis.

Despite the overall strong performance, the data reveal notable variability, particularly in the low- and medium-analyzability conditions. This dispersion may be due to individual differences such as prior exposure to classical and quantum programming, academic background, or the sequence in which tasks were presented. It is noteworthy that the minimum score in the low-analyzability group was 50\%, and 62.5\% in the medium group, suggesting that circuit complexity can significantly challenge less experienced students, even in structured experimental settings.

Although the order of circuit presentation was randomized for all participants across the four experiments, only experiments 2-BScMay and 4-BScNov had sufficiently large and balanced sample sizes to allow a meaningful analysis of how presentation order might affect results. In these two groups, participants were randomly assigned to one of several predefined sequences (e.g., Low–Medium–High, Medium–High–Low, etc.), ensuring representation of all possible orders and enabling comparisons between them. In contrast, the remaining experiments (1-MSc and 3-Prof) had smaller or more uneven sample distributions across orders, which limited the feasibility of drawing conclusions about ordering effects.

The results presented in Table~\ref{tab:descriptive_by_order_2-BScMay} illustrate how scores vary based on the order of circuit presentation in 2-BScMay. In the "Low, Medium, High" order (N = 15), the scores for low- and medium-analyzability circuits have high means (90.00\% and 90.83\%, respectively) but exhibit considerable variability (standard deviations of 17.15 and 14.54). In contrast, high-analyzability circuits consistently achieve 100\% with no variability. In the "Low, High, Medium" order (N = 19), the mean \textit{Low Score} is slightly lower (84.21\% with SD = 19.02) compared to other orders, while the \textit{Medium Score} remains similar (90.79\% with SD = 13.72), and the \textit{High Score} is fixed at 100\%. On the other hand, in the "Medium, High, Low" and "High, Medium, Low" orders (N = 16 and 13, respectively), the mean scores are around 92–94\% for the \textit{Low Score}. Approximately 91\% for the \textit{Medium Score}, with low variability, confirming that in these cases, high-analyzability circuits are consistently evaluated. Notably, in the "Medium, Low, High" order (N = 15), more significant variability is observed in the \textit{High Score} ($\overline{x}$ = 93.33\%, SD = 25.83), suggesting that the sequence of circuit presentation may influence evaluation stability, possibly due to ordering effects or participant adaptation.

\begin{table}[htb]
\centering
\caption{Descriptive statistics by circuits order for 2-BScMay scores}
\label{tab:descriptive_by_order_2-BScMay}
\begin{tabular}{lcccccccc}
\toprule
\textbf{Circuit Order} & \textbf{N} & \multicolumn{2}{c}{\textbf{Low}} & \multicolumn{2}{c}{\textbf{Medium}} & \multicolumn{2}{c}{\textbf{High}} \\
\cmidrule(lr){3-4} \cmidrule(lr){5-6} \cmidrule(lr){7-8}
 &  & $\overline{x}$ & std. dev. & $\overline{x}$ & std. dev. &$\overline{x}$ & std. dev. \\
\midrule
\textbf{Low, Medium, High} & 15  & 90.00\%  & 17.15\%  & 90.83\%  & 14.54\%  & 100.00\%  & 0.00\%  \\
\textbf{Low, High, Medium} & 19  & 84.21\%  & 19.02\%  & 90.79\%  & 13.72\%  & 100.00\%  & 0.00\%  \\
\textbf{Medium, High, Low} & 16  & 92.97\%  & 11.15\%  & 91.41\%  & 10.92\%  & 100.00\%  & 0.00\%  \\
\textbf{High, Medium, Low} & 13  & 94.23\%  & 8.26\%  & 91.35\%  & 13.86\%  & 100.00\%  & 0.00\%  \\
\textbf{Medium, Low, High} & 15  & 86.67\%  & 17.34\%  & 92.50\%  & 11.38\%  & 93.33\%  & 25.83\%  \\
\textbf{High, Low, Medium} & 6  & 89.58\%  & 12.28\%  & 87.50\%  & 15.81\%  & 100.00\%  & 0.00\%  \\
\bottomrule
\end{tabular}
\end{table}

On the other hand, Table~\ref{tab:descriptive_by_gender_2-BScMay} provides an analysis of the scores broken down by gender. In the male group (N = 70), the mean scores for low- and medium-analyzability circuits are 88.75\% and 91.43\%, respectively, with slight variability (SD of 15.67 and 12.69). Male participants obtained an average \textit{High Score} of 98.57\% with a standard deviation of 11.93, indicating some dispersion. In the female group (N = 11), scores are slightly higher in the \textit{Low Score} category ($\overline{x}$ = 90.91\%, SD = 14.88) and reach the maximum (100\%) in high-analyzability circuits, showing no variability. Meanwhile, the group categorized as "Other" (N = 3) shows a mean of 95.83\% in the \textit{Low Score} and 91.67\% in the \textit{Medium Score}, with perfect results in the \textit{High Score}. Although the sample size for the "Other" group is very small, these data suggest that, regardless of gender, high-analyzability circuits are evaluated consistently. In contrast, the scores for low- and medium-analyzability circuits exhibit moderate variations.

\begin{table}[htb]
\centering
\caption{Descriptive statistics by gender for 2-BScMay scores}
\label{tab:descriptive_by_gender_2-BScMay}
\begin{tabular}{lcccccccc}
\toprule
\textbf{Gender} & \textbf{N} & \multicolumn{2}{c}{\textbf{Low}} & \multicolumn{2}{c}{\textbf{Medium}} & \multicolumn{2}{c}{\textbf{High}} \\
\cmidrule(lr){3-4} \cmidrule(lr){5-6} \cmidrule(lr){7-8}
 &  & $\overline{x}$ & std. dev. & $\overline{x}$ & std. dev. &$\overline{x}$ & std. dev. \\
\midrule
\textbf{Male} & 70  & 88.75\%  & 15.67\%  & 91.43\%  & 12.69\%  & 98.57\%  & 11.93\%  \\
\textbf{Female} & 11  & 90.91\%  & 14.88\%  & 88.64\%  & 14.20\%  & 100.00\%  & 0.00\%  \\
\textbf{Other} & 3  & 95.83\%  & 7.22\%  & 91.67\%  & 14.43\%  & 100.00\%  & 0.00\%  \\
\bottomrule
\end{tabular}
\end{table}

\begin{figure*}[htb]
    \centering
    \begin{minipage}{0.43\textwidth}
        \centering
        \includegraphics[width=\linewidth]{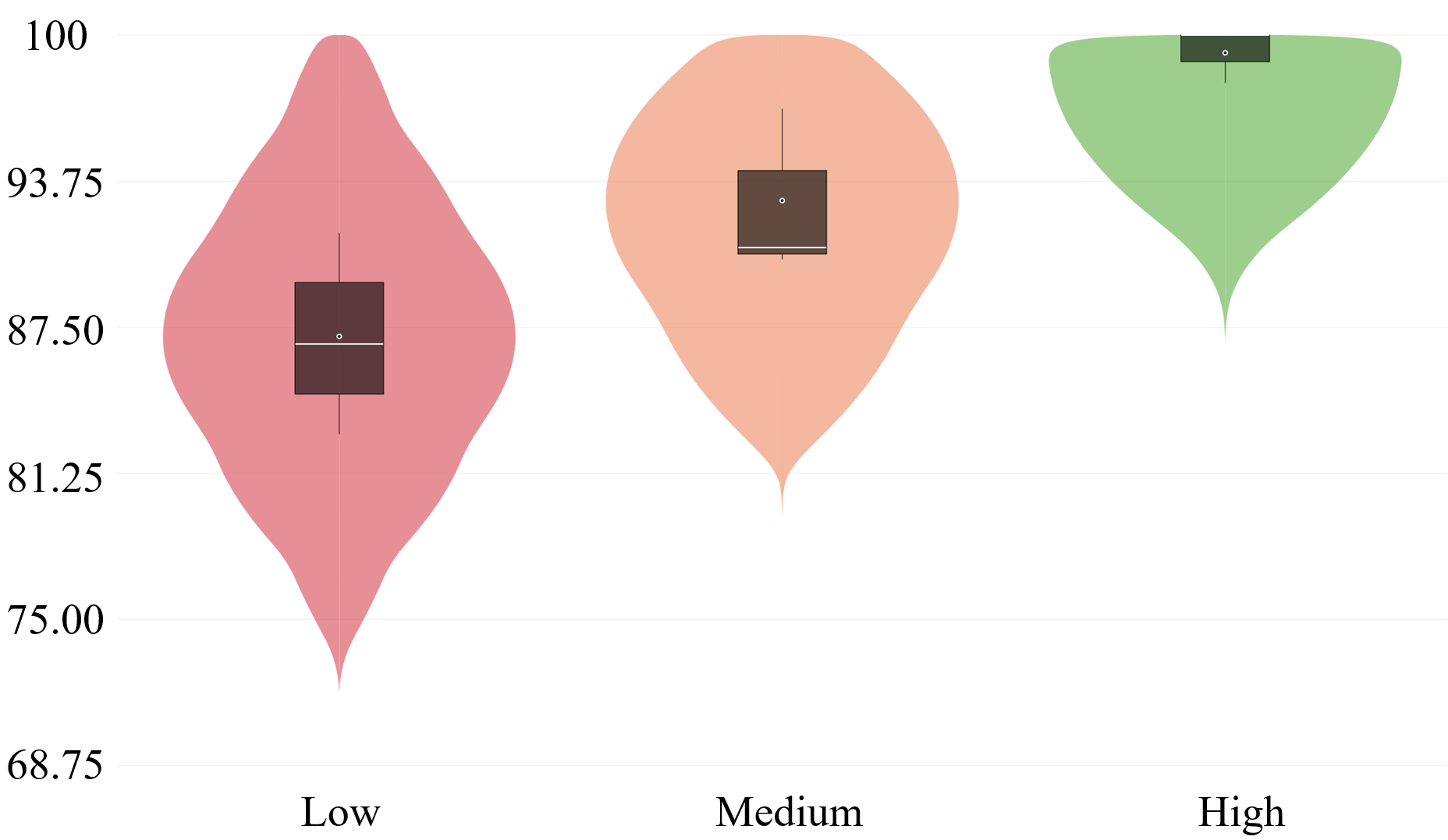}
        \subcaption{By age groups}
    \end{minipage}%
    \hspace{1cm}
    \begin{minipage}{0.43\textwidth}
        \centering
        \includegraphics[width=\linewidth]{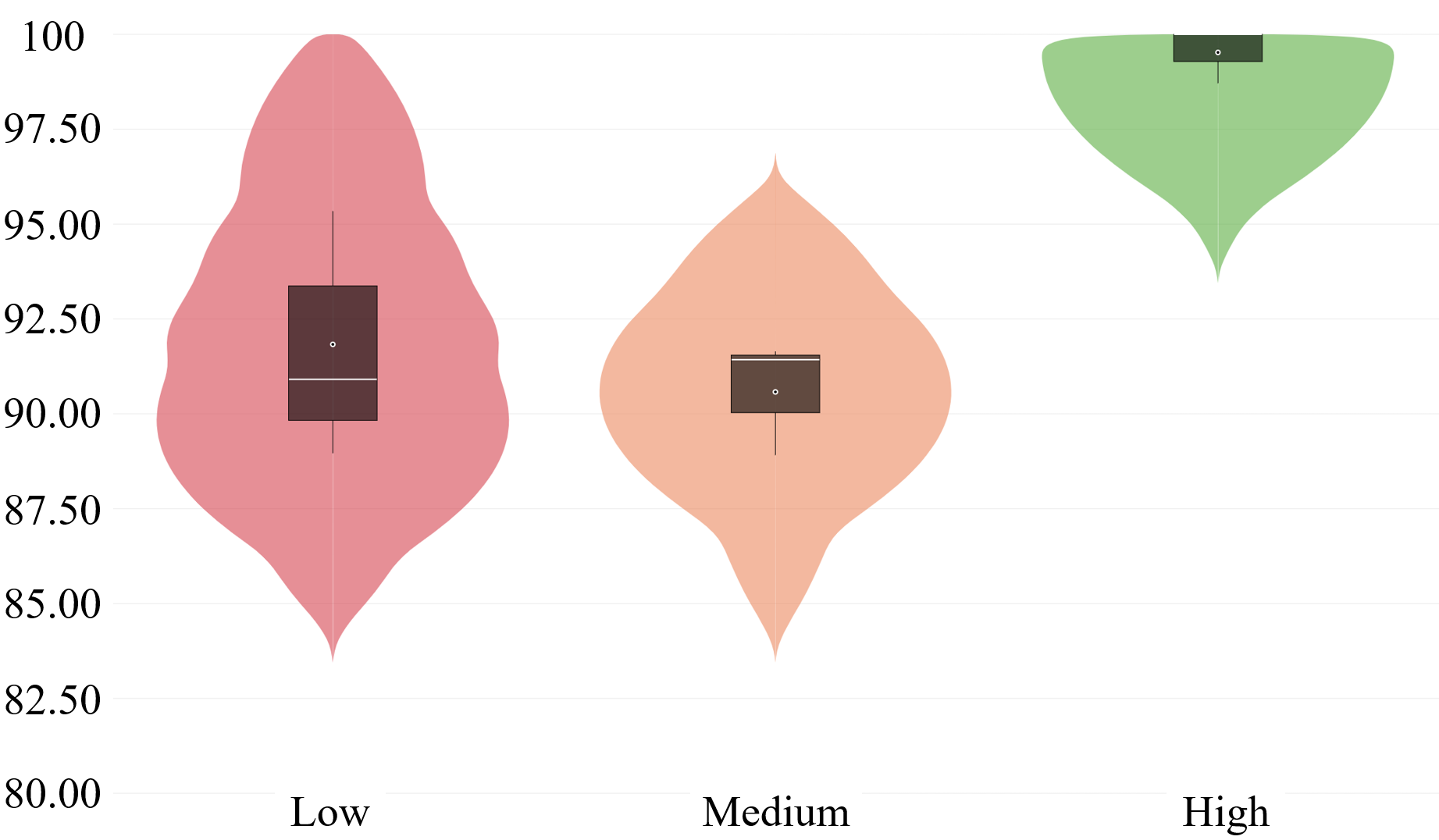}
        \subcaption{By gender}
    \end{minipage}
    
    \vspace{0.5cm}

    \begin{minipage}{0.43\textwidth}
        \centering
        \includegraphics[width=\linewidth]{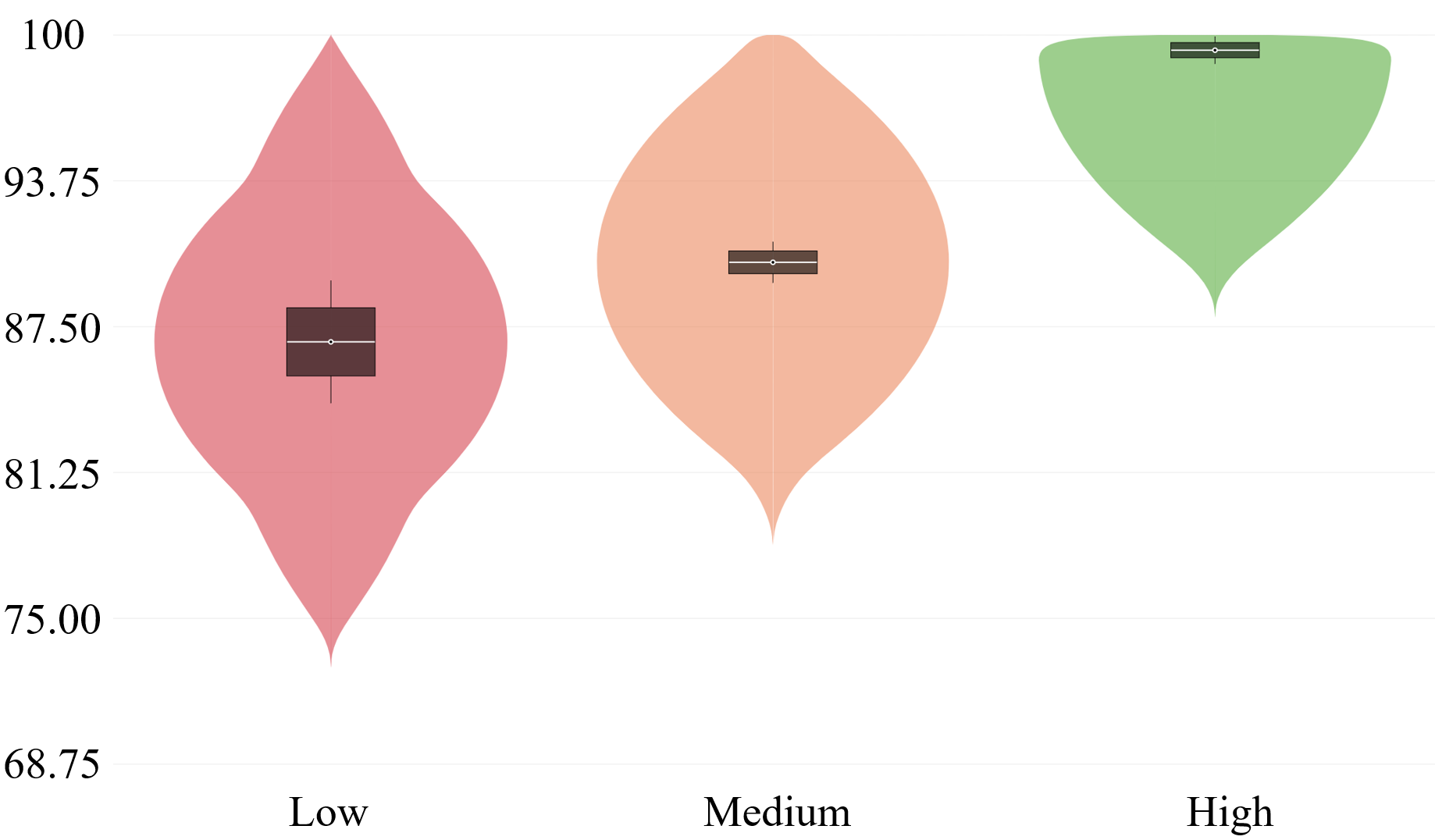}
        \subcaption{By level of education}
    \end{minipage}%
    \hspace{1cm}
    \begin{minipage}{0.43\textwidth}
        \centering
        \includegraphics[width=\linewidth]{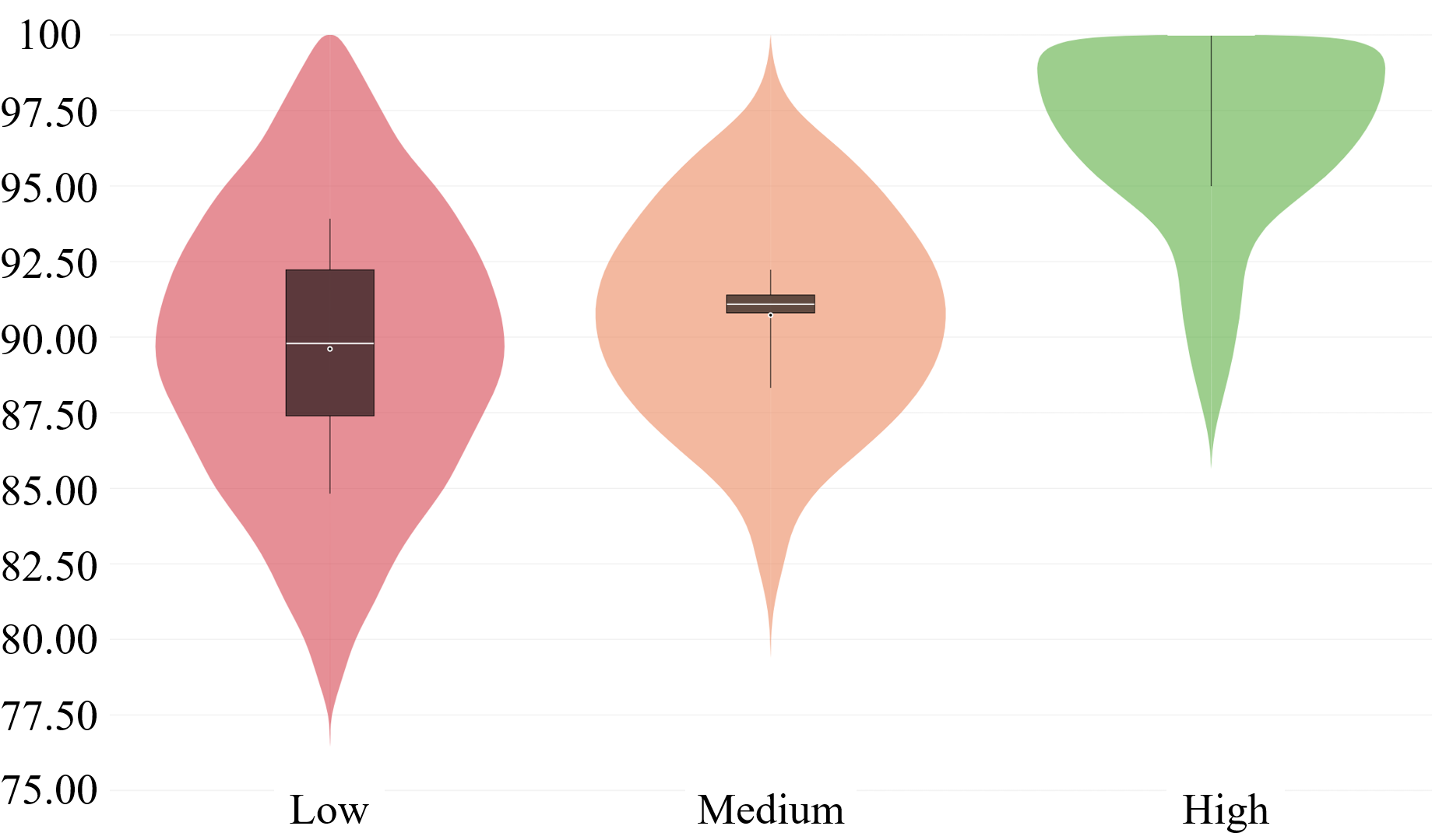}
        \subcaption{By circuits order}
    \end{minipage}

    \vspace{0.5cm}

    \begin{minipage}{0.43\textwidth}
        \centering
        \includegraphics[width=\linewidth]{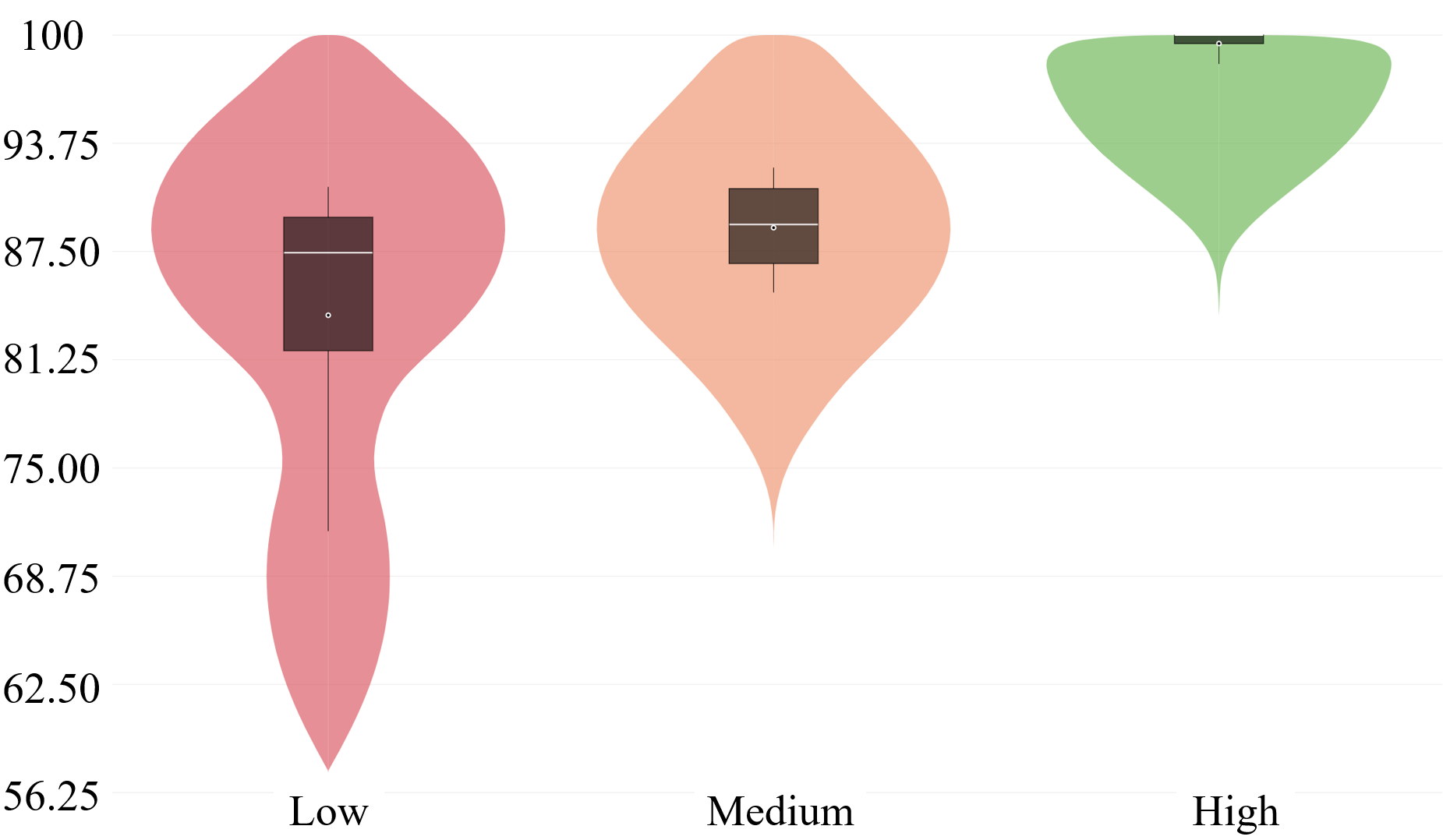}
        \subcaption{By classical computing experience}
    \end{minipage}%
    \hspace{1cm}
    \begin{minipage}{0.43\textwidth}
        \centering
        \includegraphics[width=\linewidth]{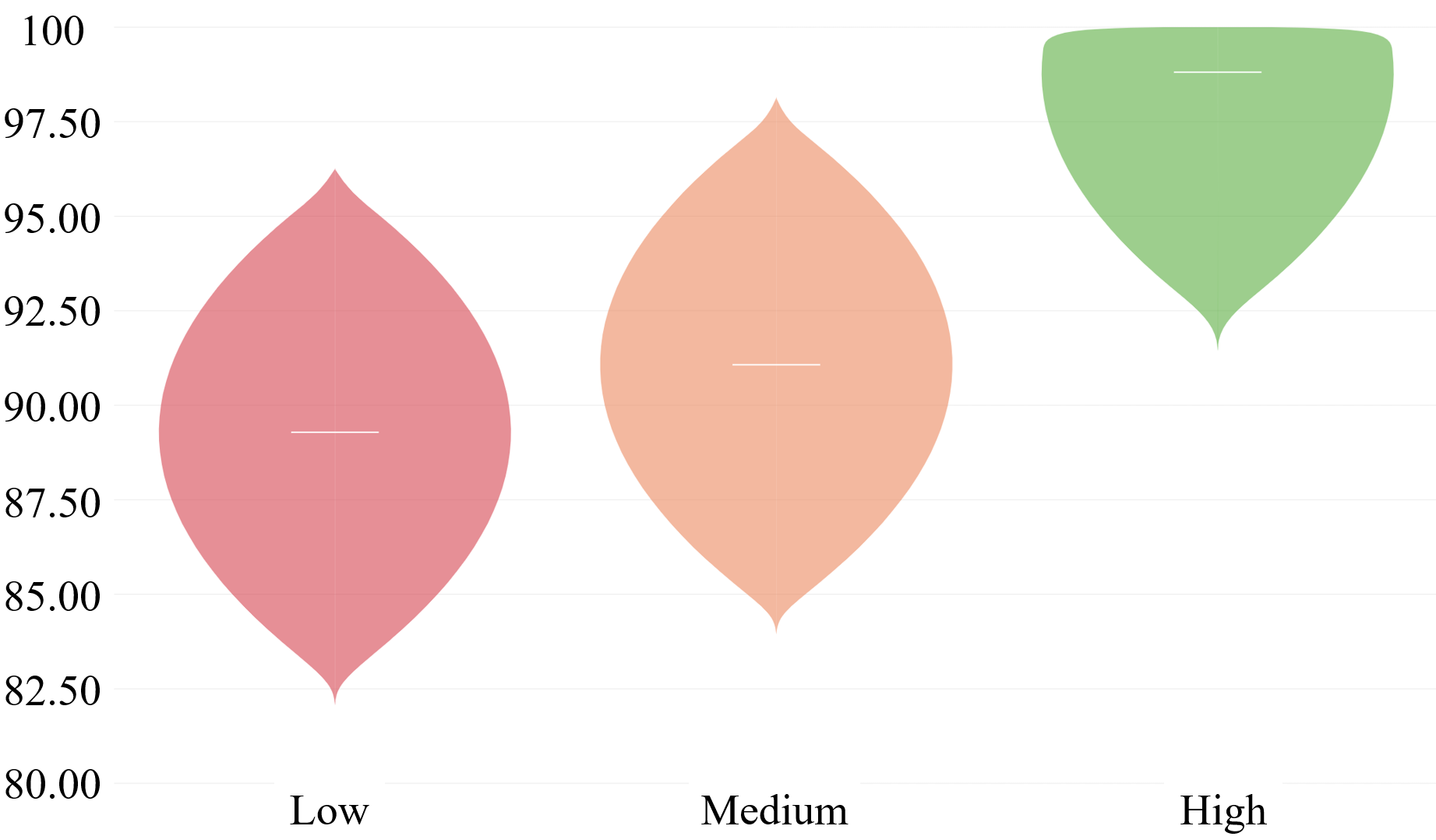}
        \subcaption{By quantum computing experience}
    \end{minipage}

    \caption{Violin plots for 2-BScMay showing the distribution of participants' scores across analyzability levels (low, medium, high) for different factors}
    \label{fig:violins_Experiment2}
\end{figure*}

Regarding the scores obtained in experiment 2-BScMay when segmented by participant characteristics, Figure~\ref{fig:violins_Experiment2} presents a set of violin plots that illustrate the distribution of results across circuits with low, medium, and high analyzability levels. Each subplot corresponds to a different factor considered in the study: (a) age group, (b) gender, (c) education level, (d) order in which the circuits were presented to the participant, (e) previous experience in classical computing, and (f) previous experience in quantum computing.

In each violin plot, the \textbf{x-axis} represents the three analyzability levels defined in the study (Low, Medium, High), while the \textbf{y-axis} shows the scores obtained by participants, normalized as a percentage from 0\% to 100\%. The shape of each violin reflects the kernel density estimation of the scores, offering a view of the distribution and concentration of values. Superimposed boxplots indicate the median and interquartile ranges, providing additional statistical insight.

This visualization enables a comparative analysis of whether the influence of analyzability on performance remains consistent across different subgroups of participants. For example, the plots allow us to observe whether participants with more quantum experience performed better in low-analyzability circuits than those with no prior background, or whether the order of circuit presentation affected the learning curve. In all six factors analyzed, the results consistently show higher median scores and reduced variability as the analyzability level increases, reinforcing the predictive validity of the proposed model.

\subsubsection{Descriptive statistical analysis of 3-Prof}

3-Prof comprises a group of participants with a more experienced and specialized profile, including both professionals and researchers with postgraduate studies (master's or doctorate) as well as some individuals with undergraduate degrees and years of experience in classical or quantum computing (Table~\ref{tab:descriptive_experiment3}). Unlike the previous experiments, prior experience in quantum computing is more common in this sample, with several participants reporting over a decade of practice in classical computing environments or sustained involvement in software quality assessment.

\begin{table}[htb]
\centering
\caption{Descriptive statistics for 3-Prof scores}
\label{tab:descriptive_experiment3}
\begin{tabular}{lcccc}
\toprule
\textbf{Quality level} & \textbf{N} & \textbf{$\overline{x}$} & \textbf{std. dev.} & \textbf{95\% CI} \\
\midrule
Low    & 15 & 89.17\%  & 18.75\%  & [79.68\%, 98.66\%] \\
Medium & 15 & 91.67\%  & 9.86\%  & [86.68\%, 96.66\%] \\
High   & 15 & 100.00\%  & 0.00\%  & [100.00\%, 100.00\%] \\
\bottomrule
\end{tabular}
\end{table}

Table~\ref{tab:descriptive_experiment3} summarizes the descriptive statistics of the scores obtained in 3-Prof, in which 15 professionals and researchers participated. In this experiment, the independent variable is the analyzability level of the quantum circuits (low, medium, and high), while the dependent variable is the average score reflecting each participant's ability to read, comprehend, and analyze the circuits.

These findings reinforce the hypothesis that greater analyzability, corresponding to enhanced structural clarity, facilitates better comprehension and analysis. Nevertheless, the variability observed in the low and medium conditions suggests that factors such as participants’ background in classical or quantum computing, familiarity with the problem domain, or cognitive strategies may influence the effectiveness with which they analyze more complex quantum circuits.

It should be noted that, due to the relatively small number of professionals in this experiment (N = 15), the randomization of the circuit presentation may not have been as rigorously controlled as desired. As a result, potential order effects and other confounding factors might have contributed to the variability observed, especially in the low and medium analyzability conditions. Future studies should aim to increase the sample size and employ more stringent randomization procedures to better isolate the impact of circuit complexity on evaluators' performance.

On the other hand, Table~\ref{tab:descriptive_by_gender_3-Prof} breaks down the results by gender. In the male group (N = 13), the mean \emph{Low Score} is 93.27\% (SD = 17.40\%), and the mean \emph{Medium Score} is 94.23\% (SD = 8.26\%), with perfect scores in high-analyzability circuits. In contrast, the female group (N = 2) shows lower means for low- (62.50\%) and medium-analyzability (75.00\%) circuits, although they also achieve 100\% in high-analyzability circuits. Although the sample size for the female group is very small, these results suggest that differences in complexity perception may exist depending on gender, with circuits of lower clarity particularly challenging for those with less experience in the field.

\begin{table}[htb]
\centering
\caption{Descriptive statistics by gender for 3-Prof scores}
\label{tab:descriptive_by_gender_3-Prof}
\begin{tabular}{lcccccccc}
\toprule
\textbf{Gender} & \textbf{N} & \multicolumn{2}{c}{\textbf{Low}} & \multicolumn{2}{c}{\textbf{Medium}} & \multicolumn{2}{c}{\textbf{High}} \\
\cmidrule(lr){3-4} \cmidrule(lr){5-6} \cmidrule(lr){7-8}
 &  & $\overline{x}$ & std. dev. & $\overline{x}$ & std. dev. &$\overline{x}$ & std. dev. \\
\midrule
\textbf{Male}   & 13 & 93.27\% & 17.40\% & 94.23\% & 8.26\% & 100.00\% & 0.00\% \\
\textbf{Female} & 2 & 62.50\% & 0.00\% & 75.00\% & 0.00\% & 100.00\% & 0.00\% \\
\bottomrule
\end{tabular}
\end{table}

Overall, these analyses indicate that the analyzability model can distinctly capture the complexity of quantum circuits, as structural clarity is consistently associated with better evaluation results. The variability observed in low- and medium-analyzability circuits highlights the influence of presentation order and individual factors, such as experience and possibly gender, on evaluator performance. These findings confirm the model's sensitivity in distinguishing between different levels of complexity, which is essential for its application in both professional and academic settings.

\subsubsection{Descriptive Statistical analysis of 4-BScNov}

4-BScNov presents a scenario similar to 2-BScMay, consisting of undergraduate students with very limited experience in quantum computing. However, this group is more extensive and covers a slightly broader age range. Table~\ref{tab:descriptive_experiment4} displays the statistical values of the scores based on circuit analyzability (low, medium, and high), as well as factors such as circuit order and prior experience level.

\begin{figure*}[htb]
    \centering
    \begin{minipage}{0.43\textwidth}
        \centering
        \includegraphics[width=\linewidth]{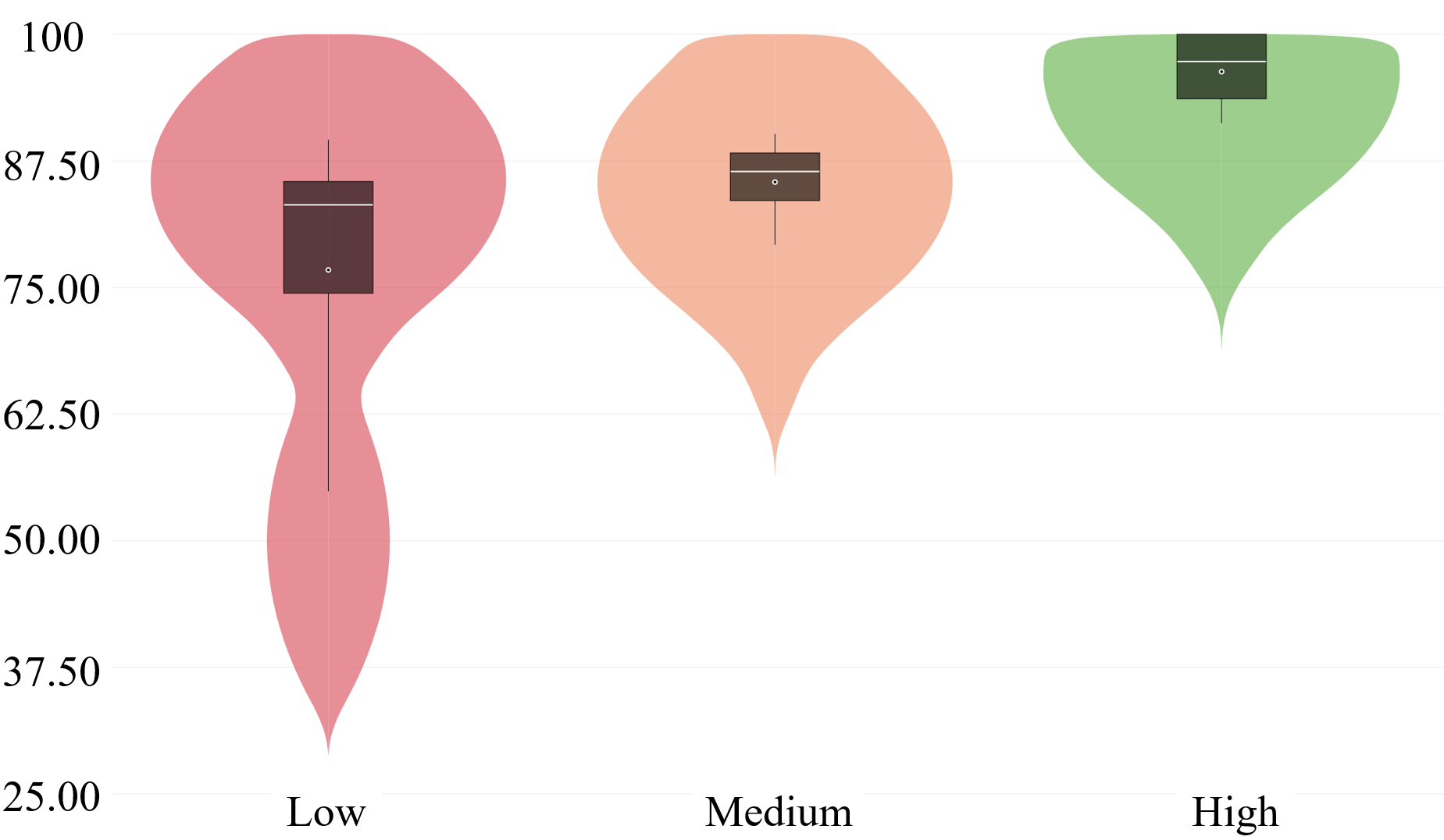}
        \subcaption{By age groups}
    \end{minipage}%
    \hspace{1cm}
    \begin{minipage}{0.43\textwidth}
        \centering
        \includegraphics[width=\linewidth]{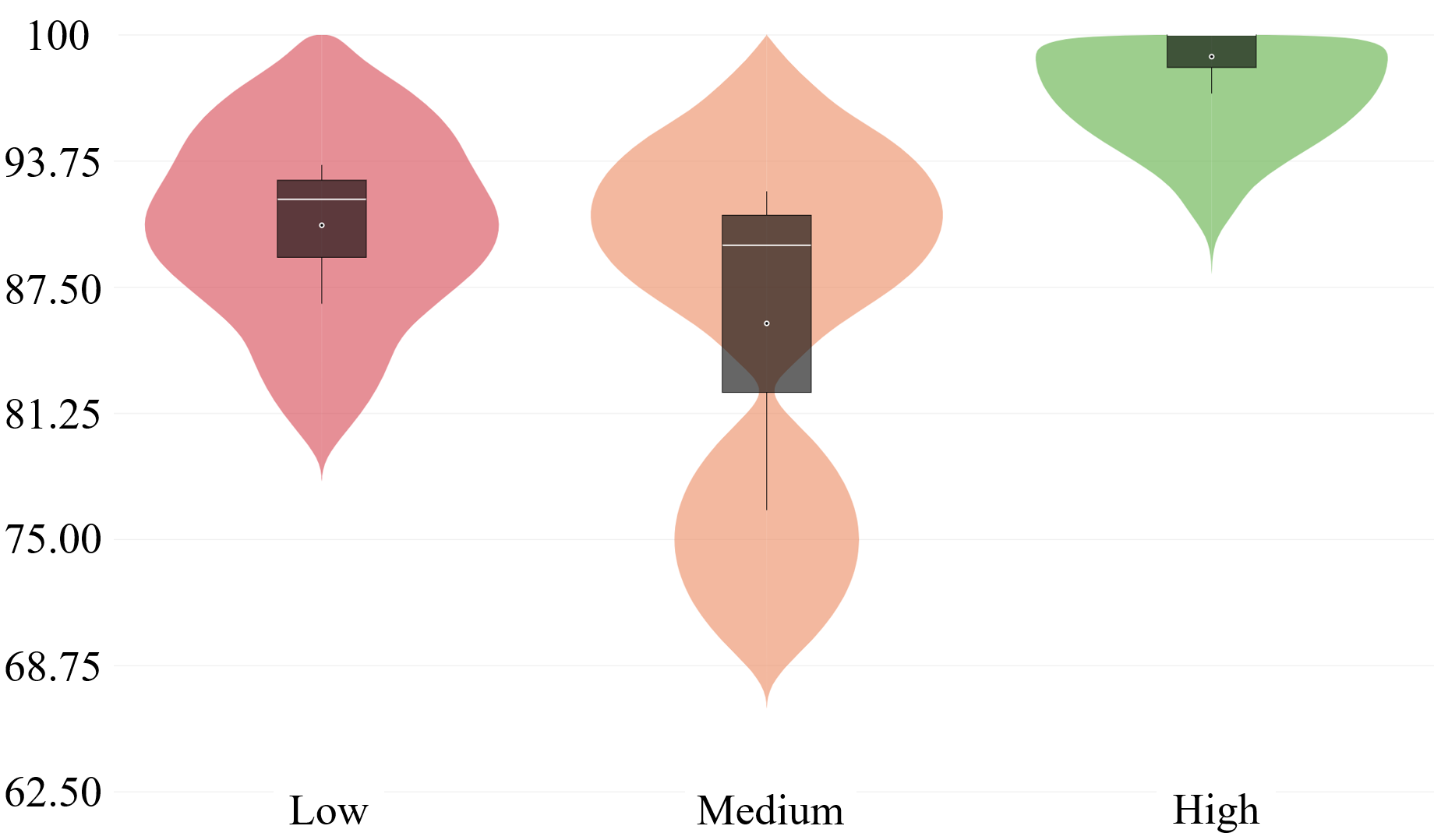}
        \subcaption{By gender}
    \end{minipage}
    
    \vspace{0.5cm}

    \begin{minipage}{0.43\textwidth}
        \centering
        \includegraphics[width=\linewidth]{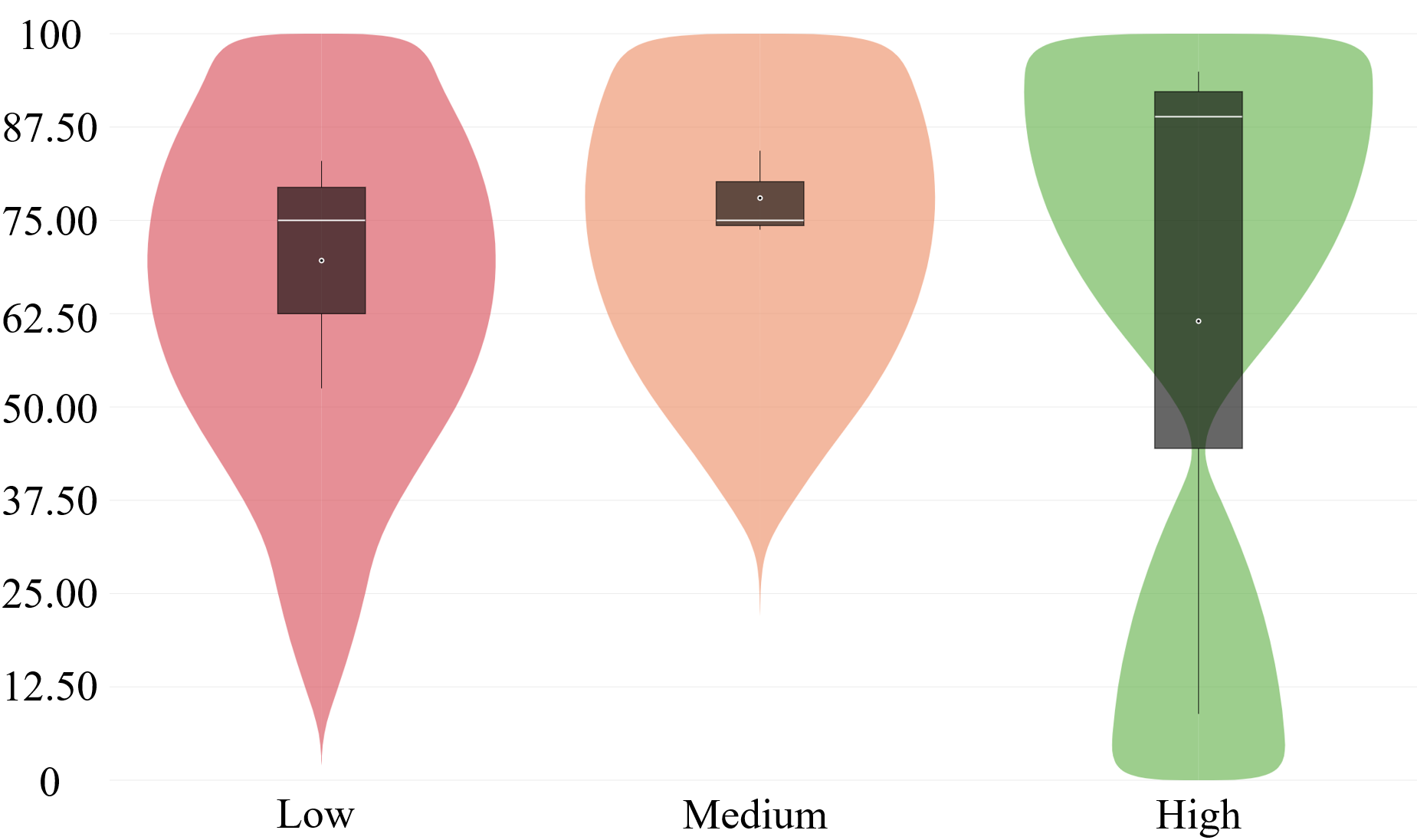}
        \subcaption{By level of education}
    \end{minipage}%
    \hspace{1cm}
    \begin{minipage}{0.43\textwidth}
        \centering
        \includegraphics[width=\linewidth]{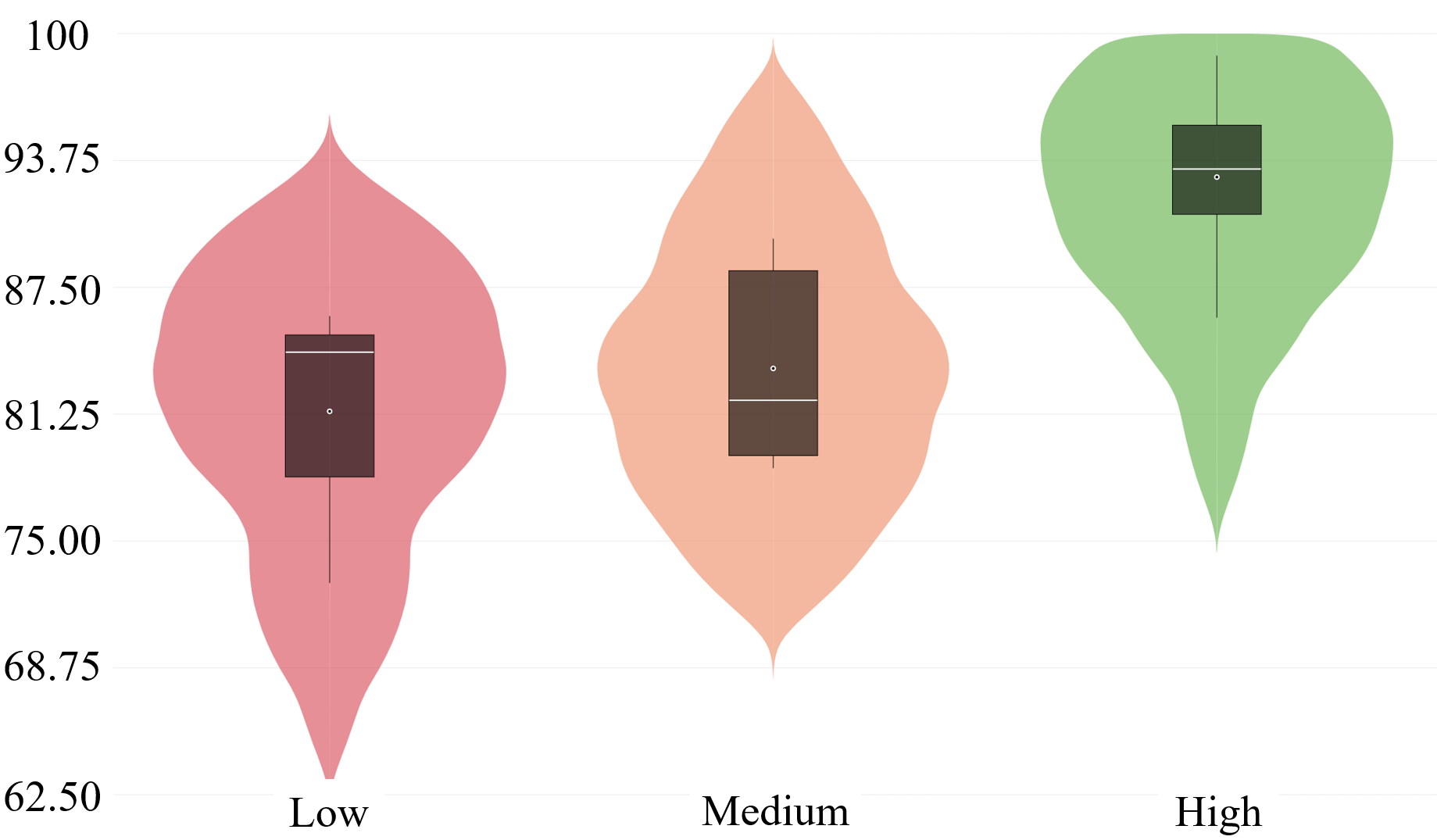}
        \subcaption{By circuits order}
    \end{minipage}

    \vspace{0.5cm}

    \begin{minipage}{0.43\textwidth}
        \centering
        \includegraphics[width=\linewidth]{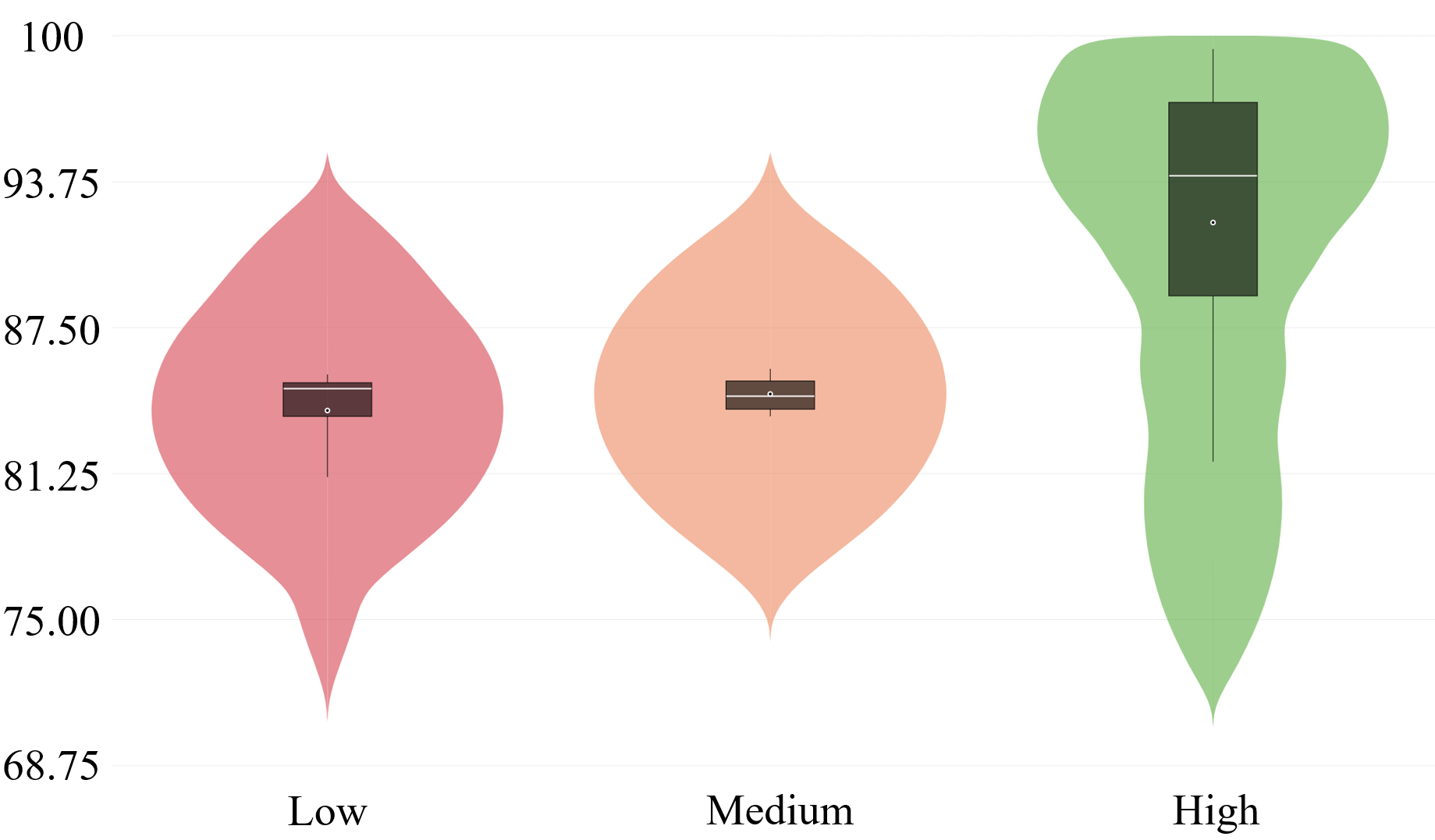}
        \subcaption{By classical computing experience}
    \end{minipage}%
    \hspace{1cm}
    \begin{minipage}{0.43\textwidth}
        \centering
        \includegraphics[width=\linewidth]{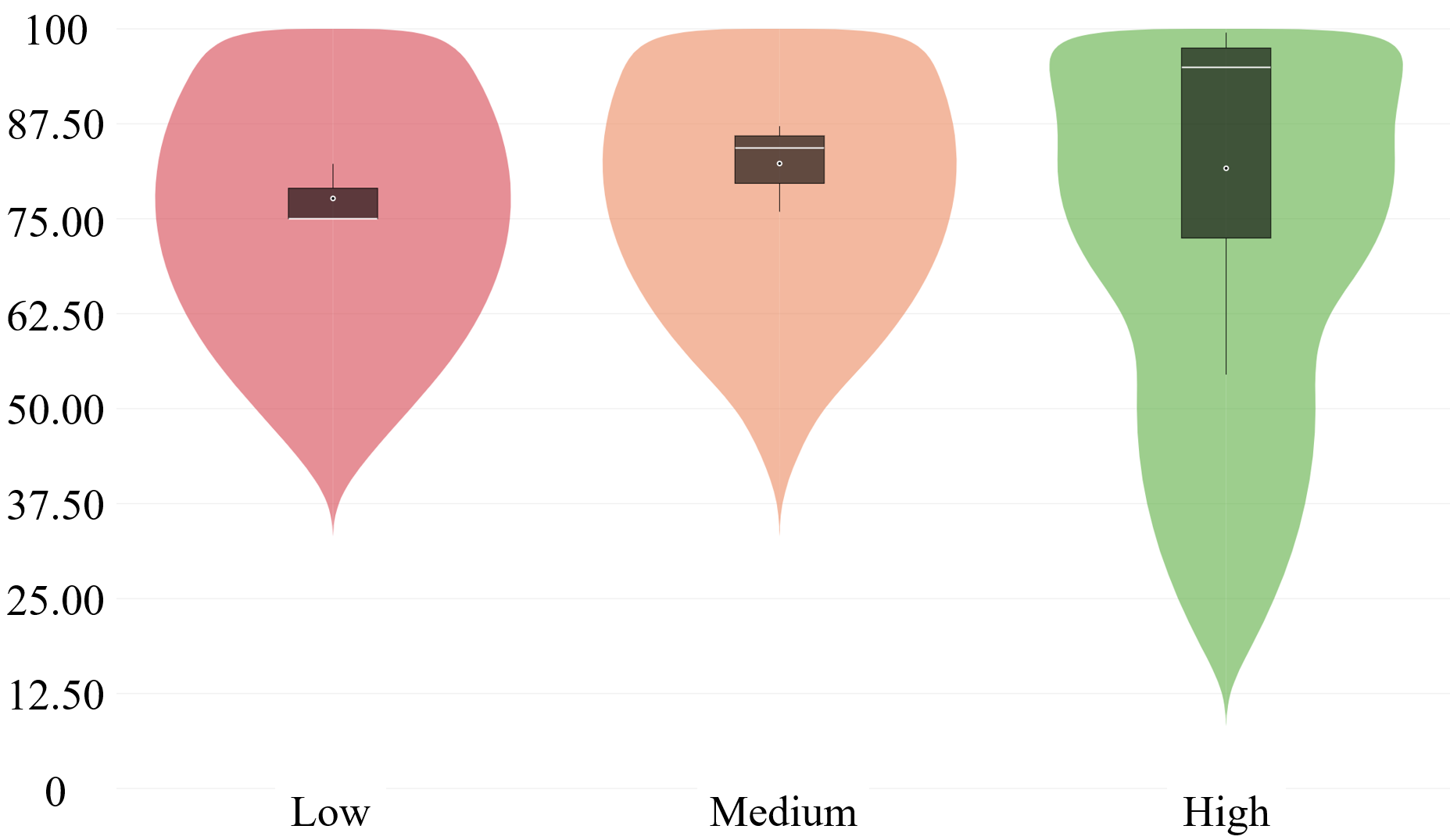}
        \subcaption{By quantum computing experience}
    \end{minipage}

    \caption{Violin plots for 4-BScNov showing the distribution of participants' scores across analyzability levels (low, medium, high) for different factors}
    \label{fig:violins_Experiment4}
\end{figure*}

Figure~\ref{fig:violins_Experiment4} presents violin plots illustrating the distribution of scores in experiment 4-BScNov, segmented by various demographic and experience-related variables. These factors include: (a) age group, (b) gender, (c) level of education, (d) order in which circuits were presented, (e) classical computing experience, and (f) quantum computing experience.

\begin{table}[htb]
\centering
\caption{Descriptive statistics for 4-BScNov scores}
\label{tab:descriptive_experiment4}
\begin{tabular}{lcccc}
\toprule
\textbf{Quality level} & \textbf{N} & \textbf{$\overline{x}$} & \textbf{std. dev.} & \textbf{95\% CI} \\
\midrule
Low    & 109 & 82.80\%  & 21.07\% & [79.14\%, 86.45\%] \\
Medium & 109 & 84.29\%  & 19.99\% & [80.89\%, 87.69\%] \\
High   & 109 & 94.15\%  & 22.75\% & [90.20\%, 98.10\%] \\
\bottomrule
\end{tabular}
\end{table}

In each plot, the \textbf{x-axis} represents the analyzability level of the circuits (Low, Medium, High), and the \textbf{y-axis} indicates the normalized score of participants from 0\% (no correct answers) to 100\% (all answers correct). The violin shapes depict the density distribution of scores within each subgroup and condition, while the internal box plots show the median and inter quartile range.

These plots reveal consistent patterns in which higher analyzability levels are associated with increased scores and reduced dispersion, regardless of the factor considered. They also provide insight into the variation between subgroups. For example, the influence of prior experience or circuit order may be visually assessed, supporting the interpretation of analyzability levels as a meaningful predictor of task performance across diverse participant profiles.

Table~\ref{tab:descriptive_experiment4} presents the descriptive statistics of the scores obtained by the 109 participants in the 4-BScNov experiment, based on the analyzability level of the quantum circuits. According to the model, circuits are classified into three levels: low analyzability (circuits with highly complex structures), medium analyzability (circuits of intermediate complexity), and high analyzability (circuits optimized to maximize clarity).

However, a broader dispersion in scores is observed in the low-analyzability condition (SD = 21.07\%), including a minimum score of 0\%. This indicates that when faced with structurally complex circuits, some participants experience significant difficulty. Even in the medium condition, although most participants perform well, variability remains noticeable. These outcomes suggest that external factors such as educational background, prior experience with classical and quantum computing, and the order of circuit presentation could influence individual performance, especially when the circuits are not highly analyzable.

In the 4-BScNov experiment, the effect of circuit order was explicitly analyzed due to the relatively large of the group, which allowed for a more balanced representation of all possible circuit sequences. This diversity enabled us to investigate whether the order of presentation influenced participant performance, as already done in the 2-BScMay experiment.

In the 4-BScNov experiment, Table~\ref{tab:descriptive_by_order_4-BScNov} shows how circuit scores vary depending on the presentation order. In the \emph{Low, Medium, High} order (N = 11), low- and medium-analyzability circuits obtain means of 76.14\% and 78.41\%, with relatively high standard deviations (30.35 and 32.15, respectively), while high-analyzability circuits reach 90.91\% (SD = 30.15\%). The \emph{Low, High, Medium} order (N = 29) shows higher means for the low and medium levels (84.91\% and 89.66\%), but also retains some variability, as reflected by the standard deviation. Similarly, in the \emph{Medium, High, Low} order (N = 25), low- and medium-analyzability circuits score around 82\% and 80.50\%. At the same time, the high level reaches 96\%, suggesting that circuits with greater structural clarity tend to yield better results. A noteworthy observation is the \emph{High, Medium, Low} order (N = 12), where the \emph{Low Score} has the lowest mean (71.88\%), potentially indicating that after first encountering a high-analyzability circuit, participants experience a noticeable contrast when faced with a low-analyzability circuit. Meanwhile, \emph{Medium, Low, High} (N = 21) reaches 100\% in high-analyzability circuits, demonstrating that when the circuit with the highest structural clarity is presented last, the transition from the medium and low levels may be more manageable. Finally, in the \emph{High, Low, Medium} order (N = 11), a mean of 84.09\% is observed for the \emph{Low Score} and 88.64\% for the \emph{Medium Score}, with some dispersion in both cases.

\begin{table}[htb]
\centering
\caption{Descriptive statistics by circuits order for 4-BScNov scores}
\label{tab:descriptive_by_order_4-BScNov}
\begin{tabular}{lcccccccc}
\toprule
\textbf{Circuit Order} & \textbf{N} & \multicolumn{2}{c}{\textbf{Low}} & \multicolumn{2}{c}{\textbf{Medium}} & \multicolumn{2}{c}{\textbf{High}} \\
\cmidrule(lr){3-4} \cmidrule(lr){5-6} \cmidrule(lr){7-8}
 &  & $\overline{x}$ & std. dev. & $\overline{x}$ & std. dev. &$\overline{x}$ & std. dev. \\
\midrule
\textbf{Low, Medium, High} & 11   & 76.14\%  & 30.35\%  & 78.41\%  & 32.15\%  & 90.91\%  & 30.15\%  \\
\textbf{Low, High, Medium} & 29   & 84.91\%  & 18.43\%  & 89.66\%  & 16.40\%  & 94.83\%  & 20.46\%  \\
\textbf{Medium, High, Low} & 25   & 82.00\%  & 21.98\%  & 80.50\%  & 22.26\%  & 96.00\%  & 20.20\%  \\
\textbf{High, Medium, Low} & 12   & 71.88\%  & 27.24\%  & 79.17\%  & 16.28\%  & 84.38\%  & 20.00\% \\
\textbf{Medium, Low, High} & 21   & 86.31\%  & 17.64\%  & 84.52\%  & 17.18\%  & 100.00\%  & 0.00\% \\
\textbf{High, Low, Medium} & 11   & 84.09\%  & 19.45\%  & 88.64\%  & 13.06\%  & 90.91\%  & 30.15\% \\
\bottomrule
\end{tabular}
\end{table}

Regarding gender distribution, Table~\ref{tab:descriptive_by_gender_4-BScNov} shows that the male group (N = 99) achieves mean scores of 83.71\% and 85.10\% in the low- and medium-analyzability levels, with standard deviations around 19, indicating considerable variability. For high-analyzability circuits, the mean rises to 93.57\%, although a standard deviation of 23.90 still suggests individual differences. In the female group (N = 9), scores in the low- and medium-analyzability circuits are somewhat lower (72.22\% and 79.17\%), with standard deviations of 32.93 and 27.95, respectively, but high-analyzability circuits are solved with a perfect score of 100\% without variation. The “Other” group (N = 1) does not allow for generalizable conclusions, although it illustrates a notable difference between the medium-analyzability circuit (50\%) and the low- and high-analyzability levels (87.5\% and 100\%, respectively). These data confirm the trend observed in other experiments: the greater structural clarity of high-analyzability circuits leads to higher scores, whereas low- and medium-analyzability levels exhibit more pronounced variability, influenced by presentation order as well as individual factors (prior experience, academic background, and possibly motivation). Overall, the validity of the analyzability model remains consistent, as it differentiates the complexity of quantum circuits while highlighting the significance of contextual and personal factors in participant performance.

\begin{table}[htb]
\centering
\caption{Descriptive statistics by gender for 4-BScNov scores}
\label{tab:descriptive_by_gender_4-BScNov}
\begin{tabular}{lcccccccc}
\toprule
\textbf{Gender} & \textbf{N} & \multicolumn{2}{c}{\textbf{Low}} & \multicolumn{2}{c}{\textbf{Medium}} & \multicolumn{2}{c}{\textbf{High}} \\
\cmidrule(lr){3-4} \cmidrule(lr){5-6} \cmidrule(lr){7-8}
 &  & $\overline{x}$ & std. dev. & $\overline{x}$ & std. dev. &$\overline{x}$ & std. dev. \\
\midrule
\textbf{Male} & 99   & 83.71\%  & 19.84\%  & 85.10\%  & 19.12\%  & 93.57\%  & 23.90\%  \\
\textbf{Female} & 9   & 72.22\%  & 32.93\%  & 79.17\%  & 27.95\%  & 100.00\%  & 0.00\%  \\
\textbf{Other} & 1   & 87.5\%  & -  & 50.00\%  & -  & 100.00\%  & -  \\
\bottomrule
\end{tabular}
\end{table}

\subsubsection{Overall summary of the results of the descriptive analysis}

\begin{figure*}[htb]
    \centering
    \begin{minipage}{0.43\textwidth}
        \centering
        \includegraphics[width=\linewidth]{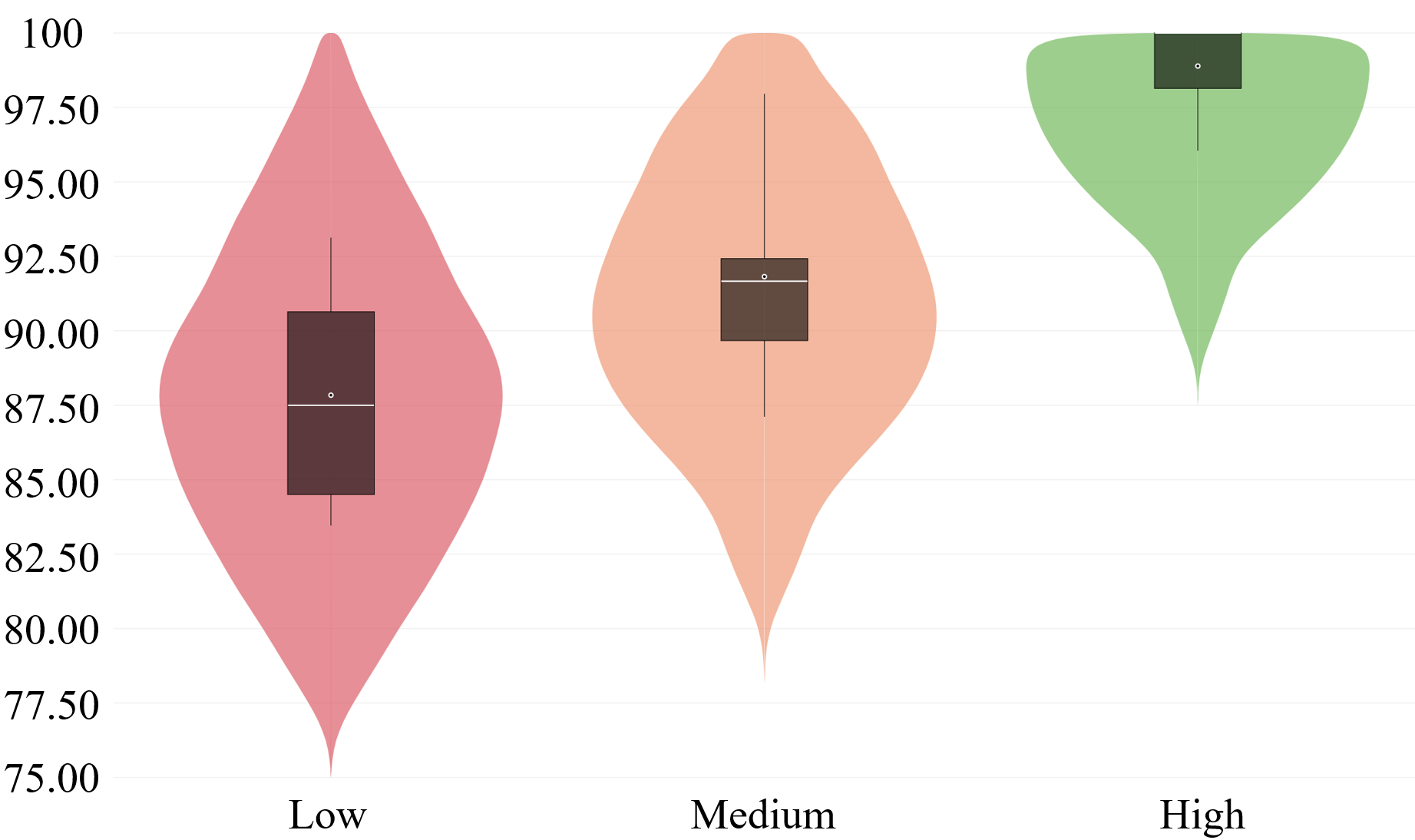}
        \subcaption{By age groups}
    \end{minipage}%
    \hspace{1cm}
    \begin{minipage}{0.43\textwidth}
        \centering
        \includegraphics[width=\linewidth]{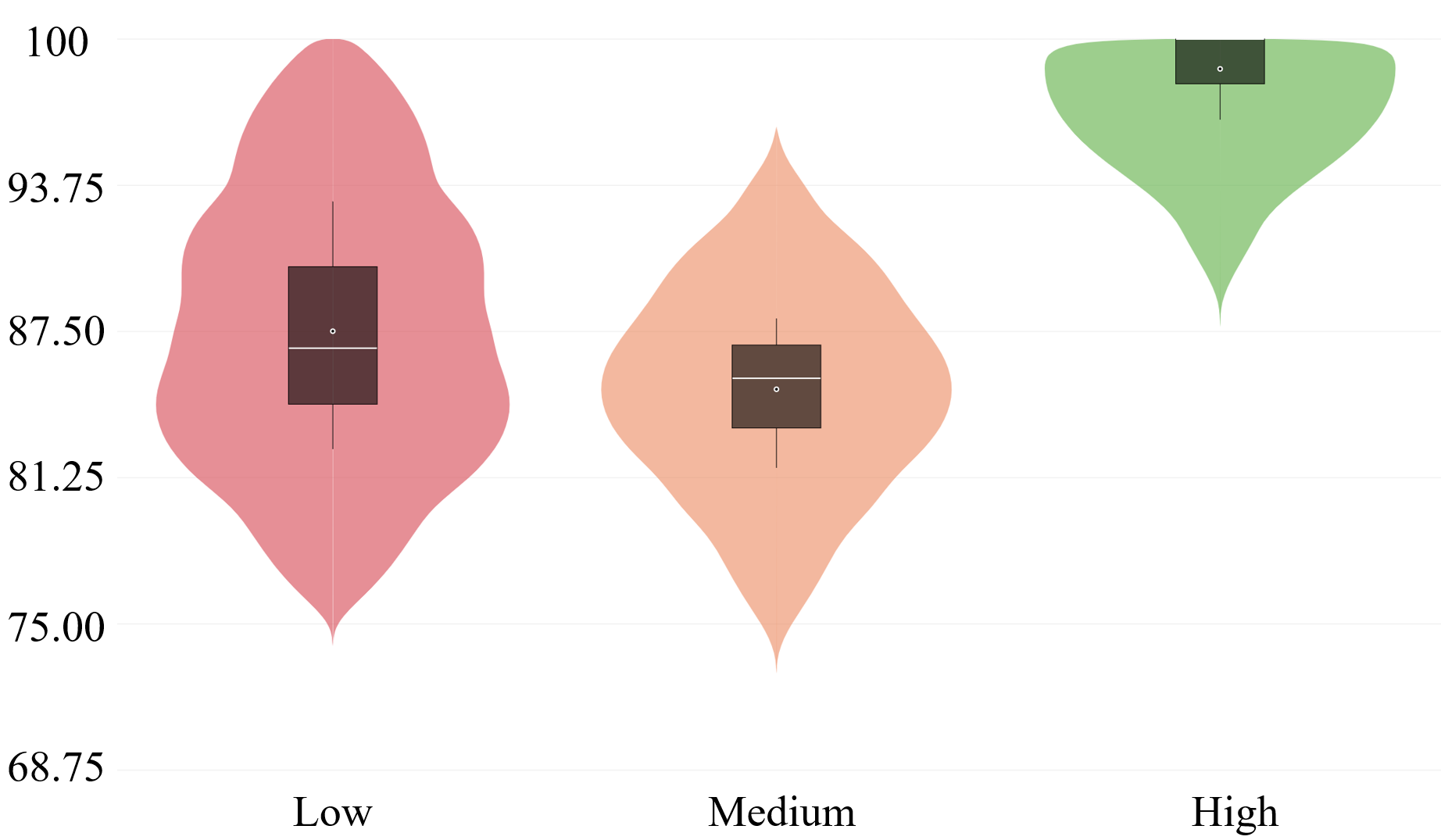}
        \subcaption{By gender}
    \end{minipage}
    
    \vspace{0.5cm}

    \begin{minipage}{0.43\textwidth}
        \centering
        \includegraphics[width=\linewidth]{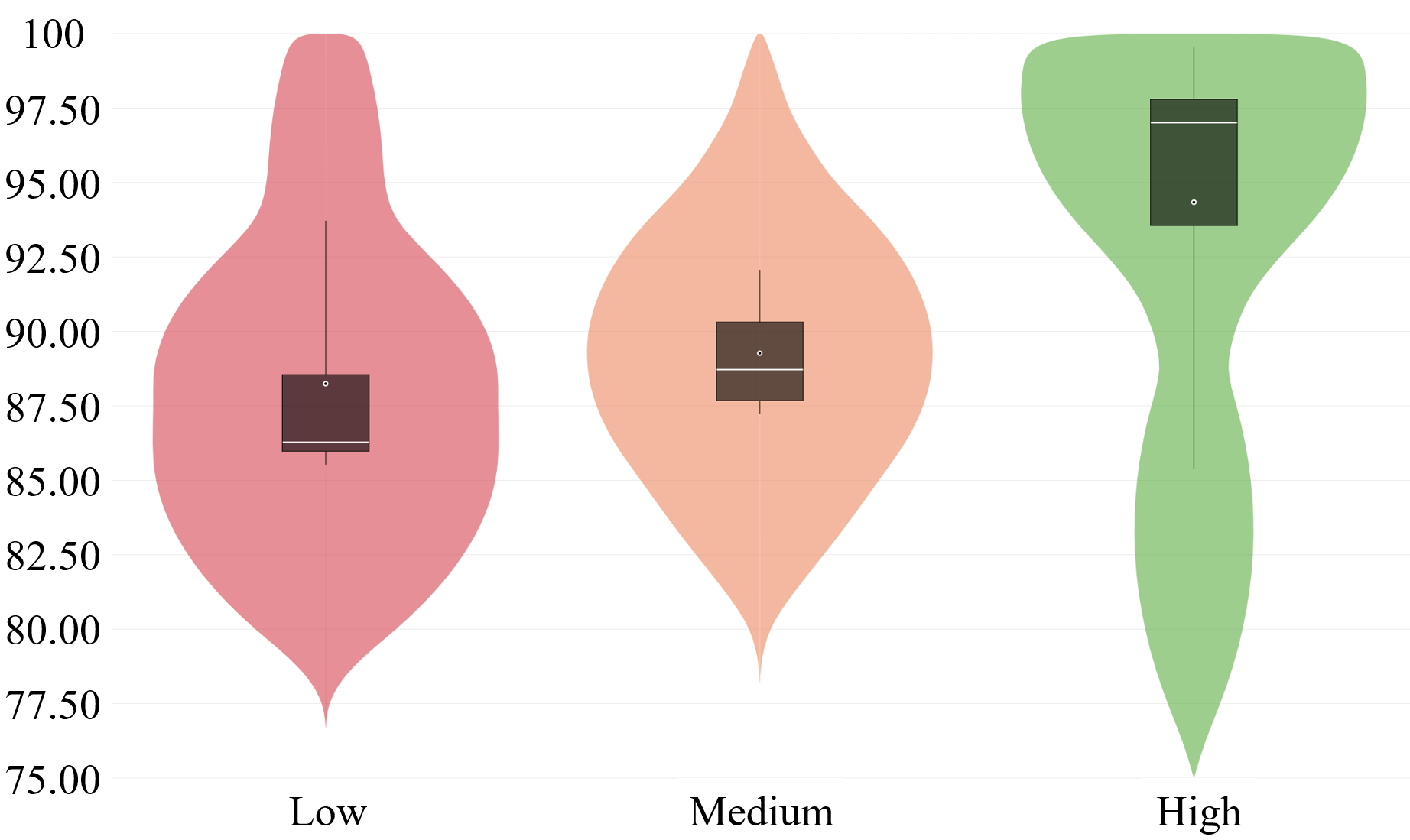}
        \subcaption{By level of education}
    \end{minipage}%
    \hspace{1cm}
    \begin{minipage}{0.43\textwidth}
        \centering
        \includegraphics[width=\linewidth]{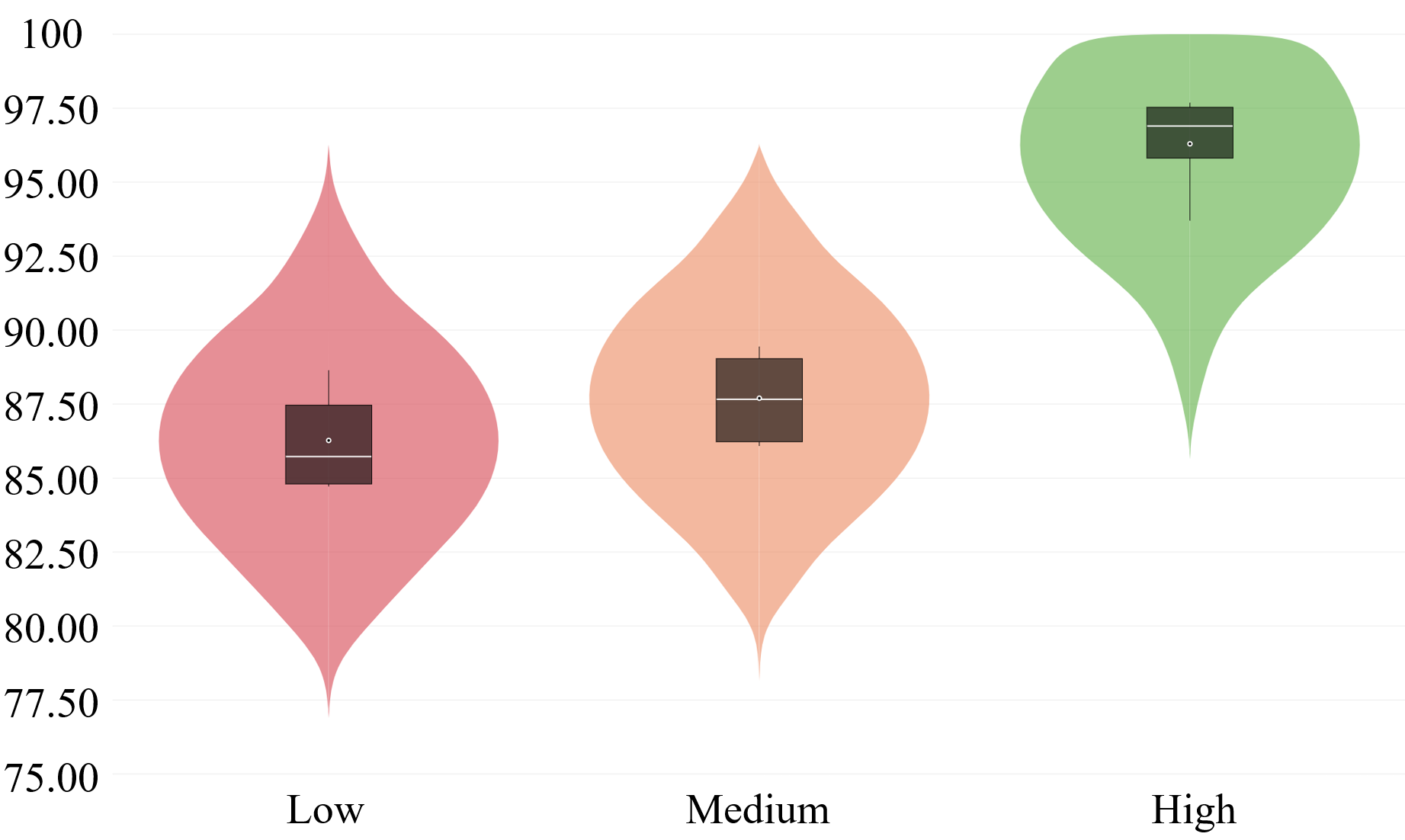}
        \subcaption{By circuits order}
    \end{minipage}

    \vspace{0.5cm}

    \begin{minipage}{0.43\textwidth}
        \centering
        \includegraphics[width=\linewidth]{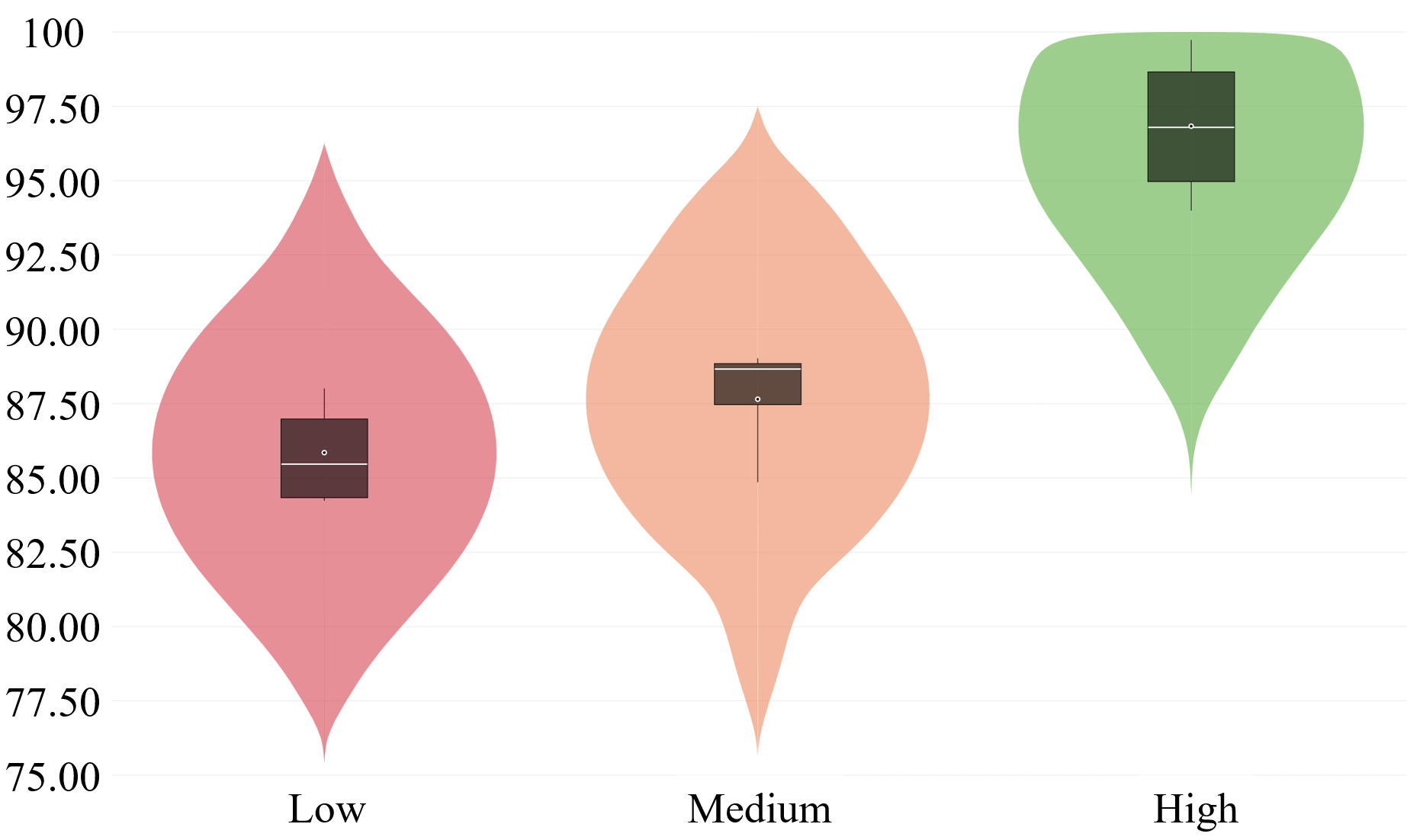}
        \subcaption{By classical computing experience}
    \end{minipage}%
    \hspace{1cm}
    \begin{minipage}{0.43\textwidth}
        \centering
        \includegraphics[width=\linewidth]{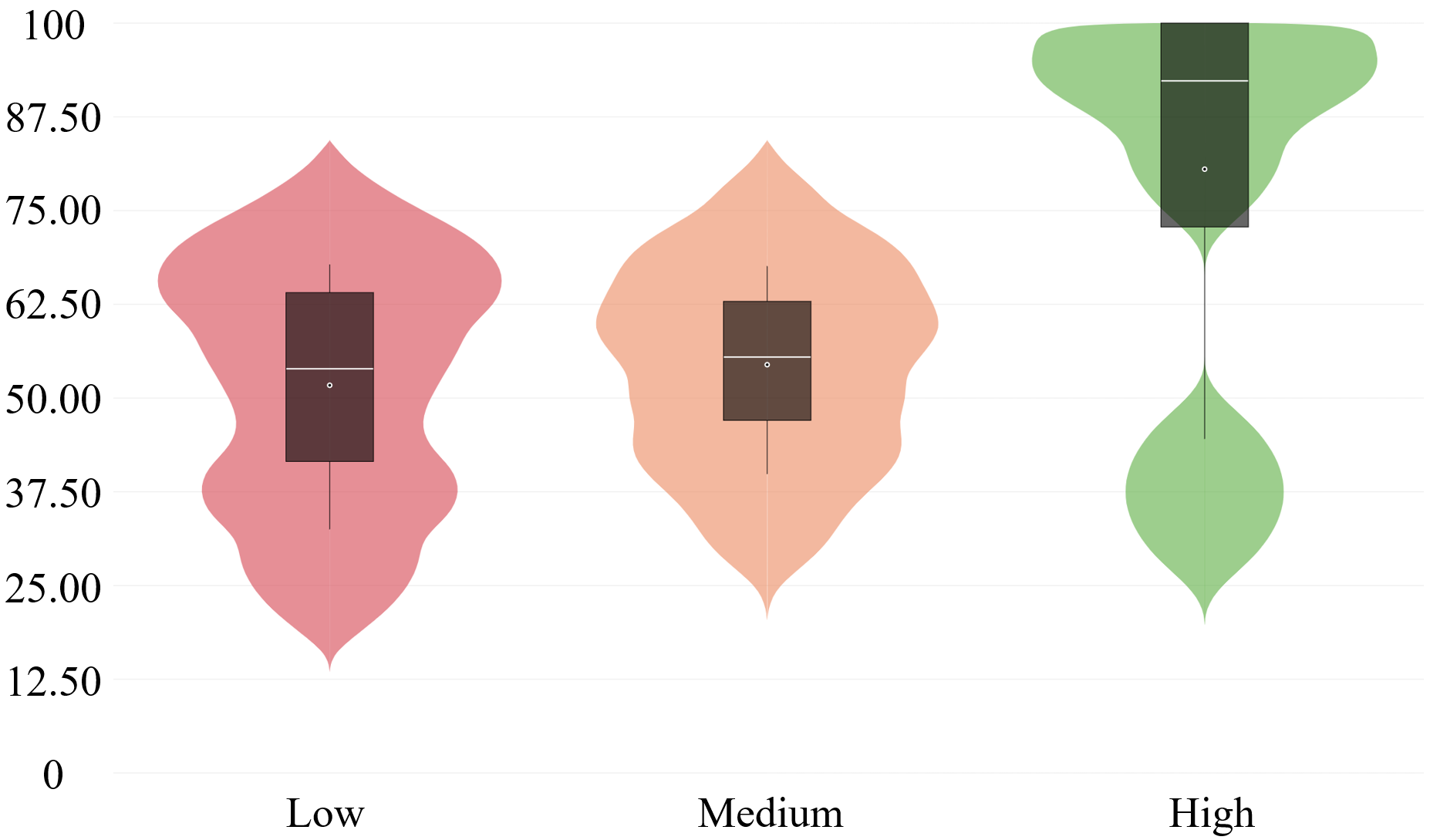}
        \subcaption{By quantum computing experience}
    \end{minipage}

    \caption{Violin plots for all experiments showing the distribution of participants' scores across analyzability levels (low, medium, high) for different factors}
    \label{fig:violins_all_experiments}
\end{figure*}

Figure~\ref{fig:violins_all_experiments} displays violin plots representing the distribution of participants' scores across all experiments, grouped by circuit analyzability level (low, medium, high) and segmented according to six key factors: age group, gender, education level, order of circuit presentation, classical computing experience, and quantum computing experience.

In each subplot, the \textbf{x-axis} categorizes circuits by their analyzability level, while the \textbf{y-axis} indicates participants’ normalized scores from 0\% to 100\%. The violin shapes show the distribution density of scores, and the embedded boxplots display the median and interquartile ranges, allowing a visual comparison of variability and central tendency.

Across all factors and experiments, a consistent trend is observed: circuits with higher analyzability levels (5/5) tend to produce higher scores with less dispersion, indicating easier comprehension. Conversely, circuits with low analyzability (1/5) result in more variable performance, especially among participants with less quantum experience. This confirms the model’s sensitivity in distinguishing circuit complexity and validates its applicability across diverse user profiles.

Interestingly, while experience and education level appear to influence variability in the more complex circuits, the \textit{order of circuit presentation} has little impact on performance. This reinforces the idea that the model's analyzability score, derived from structural metrics, is a stronger determinant of participant performance than contextual factors.

\subsection{Contrast of Hypotheses}

To determine whether there are statistically significant differences among the three analyzability levels (low, medium, and high) proposed by the model, a hypothesis test is conducted using the non-parametric Kruskal-Wallis test \cite{kruskal1952}. This test is particularly suitable for comparing more than two independent groups when the dependent variable does not follow a normal distribution, as in this study, where participants' scores in quantum circuit comprehension tasks are measured. Following this, we apply the weighted Stouffer method to compute the combined test statistic \(Z\) and the \(p\)-value, considering each experiment's sample size and concluding with the meta-analysis.

The following section details the procedure and results obtained for each experiment when performing the hypothesis test using Kruskal-Wallis.

\subsubsection{Contrast of hypotheses for 1-MSc}

As previously explained, the null hypothesis (\(H_0\)) states no statistically significant differences in participants' average scores when analyzing low-, medium-, and high-analyzability circuits. In contrast, the alternative hypothesis (\(H_a\)) expects that circuits with greater structural clarity (high analyzability) will result in higher scores. To test these hypotheses, the Kruskal-Wallis test is applied, which is suitable for comparing more than two independent groups when the dependent variable does not follow a normal distribution. In the analysis conducted, a test statistic of \(H = 3.3946\) and a \(p\)-value of 0.1832 were obtained. Since the \(p\)-value exceeds the significance threshold of 0.05, the null hypothesis is not rejected, indicating that, in this group of master's students, differences in the analyzability level of quantum circuits do not result in significant differences in the scores obtained.

The small sample size (\(N = 14\)) and the homogeneous yet slight technical knowledge of the group may have limited the magnitude of the observed differences between analyzability levels. All participants in the 1-MSc group shared a similar academic background but great limitations on Quantum Computing. We believe this might have led to an heterogeneous performance while trying to perform the experimental tasks.

\subsubsection{Contrast of hypotheses for 2-BScMay}

In 2-BScMay, which involved undergraduate students, the Kruskal-Wallis test was conducted to assess whether there were significant differences in the scores obtained for low-, medium-, and high-analyzability circuits, as defined by the model. The analysis resulted in a test statistic of \(H = 43.4129\) and a \(p\)-value less than 0.0001, allowing the rejection of the null hypothesis at the established significance level (\(\alpha = 0.05\)). This result indicates that the differences in scores across the three analyzability levels are statistically significant.

The model effectively differentiates between circuits of varying structural clarity. Specifically, circuits classified as highly analyzable tend to produce better participant evaluation results, supporting the alternative hypothesis that the analyzability level determines the ease of understanding and analyzing circuits. It is worth noting that, in this experiment, variability in participants' experience and educational level may have influenced the magnitude of the observed differences, reinforcing the importance of considering external factors in the analysis.

In conclusion, the findings from 2-BScMay confirm the hypothesis that the \textbf{analyzability level} significantly influences participants' scores. While high-analyzability circuits are generally more accessible to most students, low-analyzability circuits reveal more pronounced differences, particularly among those with less experience in classical computing or a lower educational level. This observation supports the \emph{Alternative Hypothesis} (\(H_a\)). It reinforces the model's utility in contexts where participants have mixed experience and have not yet reached an advanced academic or professional level, demonstrating that the structural clarity of quantum circuits is a key factor in their rapid comprehension and compelling analysis.

\subsubsection{Contrast of hypotheses for 3-Prof}

In 3-Prof, which involved professionals and researchers with high specialization in quantum computing and software quality, the Kruskal-Wallis test was applied to determine whether there were significant differences in the scores obtained based on the analyzability level of the circuits (low, medium, and high). The analysis yielded a test statistic of \(H = 8.0734\) and a \(p\)-value of 0.0177, allowing the rejection of the null hypothesis at the 0.05 significance level. These results indicate statistically significant differences among the groups of circuits according to their analyzability level. That is, even within a group of subjects experienced, scores vary significantly depending on the clarity and structure of the circuits; circuits classified as highly analyzable are associated with better performance in comprehension and analysis, whereas those with low analyzability present difficulties reflected in lower scores.

Overall, these results confirm the hypothesis that higher analyzability translates into better results, as evidenced by clearer circuits enabling optimal performance (100\%), while those with lower clarity exhibit a broader score dispersion. However, the variability observed in the low- and medium-analyzability conditions highlights the influence of external factors. This suggests the need to consider these elements for a more comprehensive and contextualized evaluation of the model in professional environments. The model’s ability to differentiate between different levels of complexity, even within a group with extensive experience and advanced training, reinforces its validity and utility in professional settings, underscoring the importance of applying structural clarity criteria to enhance the design and evaluation of circuits in quantum and hybrid systems.

As in the 1-MSc group happened, the sample size of this group is quit small and similar results might be expected in both groups. However, although the 3-Prof sample is heterogeneous in terms of professional profiles (researchers, developers, quality analysts), educational paths, and years of experience, almost all the participants in this group shared a deeper knowledge on Quantum Computing. This fact might have led to a better performance in the experimental tasks and, thus, the capability for rejecting the null hypothesis.

\subsubsection{Contrast of hypotheses for 4-BScNov}

In the 4-BScNov experiment, the Kruskal-Wallis test was conducted to determine whether the scores obtained by participants for low-, medium-, and high-analyzability circuits differed significantly. The analysis yielded a test statistic of \(H = 58.3928\) and a \(p\)-value of \(p < 0.0001\), allowing the rejection of the null hypothesis at the conventional significance level (\(\alpha = 0.05\)). These results provide strong evidence of statistically significant differences between the analyzability levels, reinforcing the hypothesis that circuits with greater structural clarity facilitate comprehension and analysis by the evaluators.

In this group of undergraduate students, whose experience in classical computing varied and who had little to no familiarity with quantum computing, broader score dispersion was observed for low-analyzability circuits, whereas circuits classified with high analyzability tended to produce more homogeneous and higher scores. This suggests that the complexity of low-analyzability circuits poses a real obstacle for individuals who have not yet achieved proficiency in quantum programming, whereas the more precise structure of high-analyzability circuits results in more consistent performance.

These findings support the hypothesis that a higher level of analyzability is associated with better results, as circuits designed to maximize structural clarity enable participants to achieve higher and more consistent scores. However, the variability observed in low- and medium-analyzability circuits underscores the importance of considering external factors and individual evaluator characteristics. Overall, these results confirm the validity of the proposed model in distinguishing between circuits of different complexity levels and support its applicability in educational settings, demonstrating that, although outliers exist, the general trend is that greater clarity in circuit design leads to improved evaluation performance.

\subsubsection{Meta-analysis results}
\label{meta-analisis_ponderado}

In addition to individually analyzing each experiment using the Kruskal-Wallis test, a meta-analysis was conducted that integrated the information from all participants, weighting the results according to the sample size of each experiment. The Kruskal-Wallis test is justified because the data, not following a normal distribution and having heterogeneous sample sizes, require a non-parametric method capable of simultaneously comparing three or more independent groups. In this regard, the Kruskal-Wallis test has enabled the evaluation of whether differences in circuit scores, classified according to their analyzability level (low, medium, and high), are statistically significant in each individual experiment.

This aggregated analysis not only reinforces the global significance of the analyzability levels but also helps address the statistical limitations of smaller groups such as 1-MSc and 3-Prof. In these cases, the limited sample size may reduce the statistical power to detect moderate effects. By combining results across experiments using weighted methods, the meta-analysis provides a more robust and reliable interpretation of the overall impact of analyzability on participants’ performance.

Table~\ref{tab:kruskal_results} summarizes the results obtained for each experiment: in experiment 1-MSc (\(N = 14\)), a test statistic of 3.3946 with a \(p\)-value of 0.1832 was obtained, indicating that, in this small and highly experienced group, the differences were not significant. In contrast, in experiment 2-BScMay (\(N = 84\)), a test statistic of 43.4129 with a \(p\)-value of 0.0001 was obtained, in experiment 3-Prof (\(N = 15\)), a test statistic of 8.0734 with a \(p\)-value of 0.0177 was found, and in experiment 4-BScNov (\(N = 109\)), a test statistic of 58.3928 with a \(p\)-value of 0.0001 was observed, evidencing significant differences in these experiments.

To complement the statistical significance analysis, epsilon squared (\( \varepsilon^2 \)) was computed as a non-parametric measure of effect size for the Kruskal–Wallis tests. This coefficient estimates the proportion of variance in the dependent variable (scores) that can be attributed to the independent factor (in this case, analyzability level). Values of 
\( \varepsilon^2 \) range from 0 to 1, where higher values indicate a greater magnitude of effect. As shown in Table~\ref{tab:kruskal_results}, all experiments except 1-MSc yielded medium to large effect sizes, reinforcing the impact of the analyzability level on participants’ circuit comprehension performance.

\begin{table}[htb]
\centering
\caption{Kruskal-Wallis test results by experiment}
\label{tab:kruskal_results}
\begin{tabular}{lcccc}
\toprule
\textbf{Experiment}       & \textbf{N} & \textbf{\textit{H}} & \textbf{\(p\)-value} & \textbf{\( \varepsilon^2 \)} \\
\midrule
1-MSc          & 14         & 3.3946     & 0.1832     & 0.2611 \\
2-BScMay       & 84         & 43.4129    & $<0.0001$  & 0.5079 \\
3-Prof         & 15         & 8.0734     & 0.0177     & 0.5384 \\
4-BScNov       & 109        & 58.3928    & $<0.0001$  & 0.5443 \\
\bottomrule
\end{tabular}
\end{table}

Given that the sample sizes and participant characteristics vary considerably across experiments, the \(p\)-values were combined using the weighted Stouffer method \cite{Stouffer1949}. This method allows integrating the results of individual tests by accounting for the weight of each sample (measured through \(\sqrt{n}\)), thus generating a global estimate of the effect. Table~\ref{tab:stouffer_results} presents the obtained values: the weighted sum of the \(z\)-scores was 84.4706, while the sum of the square roots of the sample sizes was 14.8997, resulting in a combined \(Z\)-score of 5.6737 and a combined \(p\)-value of \(2.73 \times 10^{-19}\).

\begin{table}[htb]
    \centering
    \caption{Results of the weighted Stouffer method}
    \label{tab:stouffer_results}
    \sisetup{
      round-mode          = places, 
      round-precision     = 4, 
    }
    \begin{tabular}{lccccc}
        \toprule
        \textbf{Experiment} & \textbf{$n$} & \textbf{$p$-value} & \textbf{$z$-score} & \textbf{$\sqrt{n}$} & \textbf{$z\,\sqrt{n}$} \\
        \midrule
        1-MSc    & 14   & 0.1832  & 0.9050  & 3.7417  & 3.3885  \\
        2-BScMay & 84   & 0.0001  & 3.7190  & 9.1652  & 34.0894 \\
        3-Prof   & 15   & 0.0177  & 2.1100  & 3.8730  & 8.1780  \\
        4-BScNov & 109  & 0.0001  & 3.7190  & 10.4403 & 38.8147 \\
        \midrule
        \multicolumn{5}{r}{\textbf{Sum of $n$}} & 222 \\
        \multicolumn{5}{r}{\textbf{$\sqrt{\sum n}$}} & 14.8997 \\
        \multicolumn{5}{r}{\textbf{$\sum (z\,\sqrt{n})$}} & 84.4706 \\
        \multicolumn{5}{r}{\textbf{Combined $Z$}} & 5.6737 \\
        \multicolumn{5}{r}{\textbf{Combined $p$-value}} & $7.3\times10^{-9}$ \\
        \bottomrule
    \end{tabular}
\end{table}

This overall result strongly indicates that, when considering the entirety of the data from the four experiments, there are statistically significant differences in the scores according to the level of circuit analyzability. In other words, the proposed model effectively differentiates between circuits with varying structural clarity, with high-analyzability circuits consistently enabling better participant performance.

The model's effectiveness is evident despite differences in participant profiles: while Experiment 1-MSc, consisting of a small and more experienced group, did not show significant differences on its own, overall the experiments with larger sample sizes (Experiments 2-BScMay and 4-BScNov) did reveal significant distinctions. The integration using the Stouffer method, which weights results based on sample size, reinforces the conclusion that the overall effect is robust.

These findings support the alternative hypothesis and empirically validate the analyzability model, demonstrating that greater structural clarity and simplicity in quantum circuits translate into superior evaluator performance. The empirical evidence reveals that the model's discriminatory power remains consistent across samples with different levels of experience, education, and motivation, study's validity for its adoption as a tool for assessing and improving quality in quantum systems. These results have significant implications for applying the model in both academic and industrial environments and establish a solid basis for future research to expand and diversify the participant sample.

\section{Discussion}
\label{discussion}

In this section, the results obtained from the empirical validation of the quantum metrics in the analyzability model are discussed in depth. The research questions formulated are analyzed and addressed, the implications of the experimental findings are interpreted based on the variables and factors considered, and the limitations and threats to the study's validity are evaluated.

\subsection{RQ1: Relation between analyzability levels and complexity perceived}

The RQ1 stated: \textit{Is there a significant relationship between the analyzability levels obtained through the proposed model and the perceived complexity of quantum software by quantum software developers?}

The empirical results obtained from the model validation indicate that, in general, there is a significant correlation between the analyzability levels computed using the model and the perceived complexity of quantum software by the participants. In experiments with sufficiently large samples, such as 2-BScMay and 4-BScNov, significant differences were found in the scores of circuits classified as low, medium, and high analyzability. This suggests that as the structural clarity of the circuit increases (i.e., moving toward circuits with high analyzability), participants tend to achieve better results in their evaluations. For instance, in 2-BScMay, the Kruskal-Wallis test yielded a high test statistic and a \(p\)-value significantly below 0.05, demonstrating that the scores between circuits of different levels are statistically significant. Similar results were observed in 4-BScNov, where variability in the scores of circuits with low and medium analyzability translated into substantial differences compared to circuits with high analyzability.

Although no significant differences were observed in 1-MSc, consisting of a small and highly experienced group, and differences were found in 3-Prof, all data integration through a weighted meta-analysis confirmed that the null hypothesis is rejected overall. This supports the premise that the proposed model effectively captures variations in the complexity perceived by evaluators, as circuits with higher analyzability consistently lead to better performance. Consequently, we conclude that there is a significant correlation between the analyzability level defined by the model and the perceived complexity of quantum software, validating the model's applicability in quantum evaluation contexts.

\subsection{RQ2: Analyzability variations}

The RQ2 stated: \textit{How do analyzability results vary when evaluating quantum systems in different experimental contexts and with diverse subject profiles?}

Empirical results demonstrate that analyzability values vary significantly depending on the experimental context and the profile of the subjects. In 1-MSc, composed of master's students with extensive experience in engineering, the differences between analyzability levels were minimal, as reflected in the Kruskal-Wallis test, which did not reach statistical significance. This suggests that a highly skilled group can overcome, to some extent, the inherent difficulties of complex circuits, mitigating variability in scores. In contrast, 2-BScMay and 4-BScNov, conducted with undergraduate students, exhibited marked differences in scores across low, medium, and high analyzability circuits. In these experiments, the lack of prior experience in quantum computing and the diversity in educational backgrounds resulted in significantly lower performance and wider score dispersion in low-analyzability circuits. 

On the other hand, in 3-Prof, despite involving professionals and researchers with high specialization, significant differences were observed in the low- and medium-analyzability conditions, indicating that even in groups with advanced knowledge, the complexity of specific circuits can present a challenge. The weighted meta-analysis, which integrates the results from all experiments (see Section~\ref{meta-analisis_ponderado}), confirms that the characteristics of the evaluated group strongly influence variability in analyzability.

In summary, these findings suggest that performance in analyzability evaluation is modulated both by the intrinsic design of the circuits and by external factors (educational level, prior experience, order of presentation). This highlights the need to consider the context and profile of the participants to appropriately interpret the results and refine the model's applicability in diverse scenarios.

\subsection{RQ3: Integral quantum circuits analyzability}

The RQ3 stated: \textit{To what extent does the quantum computing-specific component of the analyzability model enable a comprehensive evaluation of analyzability in quantum circuits?}

The empirical results obtained in this study indicate that incorporating quantum computing-specific metrics is essential for achieving a comprehensive evaluation of analyzability in quantum circuits. Across the experiments, circuits classified as highly analyzable exhibited favorable quantum indicators, such as lower circuit depth, reduced quantum cyclomatic complexity, and more considerable qubit stability, which translated into high scores and a perception of low complexity among evaluators. Conversely, low-analyzability circuits showed significantly worse values in these metrics, reflecting higher intrinsic complexity and resulting in lower performance in analysis tasks. This directly proportional relationship confirms that as quantum metrics indicate a higher level of complexity (e.g., increased circuit depth or quantum cyclomatic complexity), analyzability decreases, and conversely, when these indicators are optimized, circuits become more straightforward to analyze.

Integrating these quantum indicators into the model complements classical metrics—already well-established through standards such as ISO/IEC 25010—and enhances the evaluative framework's ability to accurately correlate objective analyzability with users' subjective perception of complexity. The findings support that quantum computing-specific metrics enable a comprehensive and valid evaluation of analyzability in quantum circuits, thereby validating the effectiveness of the proposed approach for assessing quality in quantum systems.

\subsection{Limitations and threats to validity}

This section discusses the potential limitations and threats to this study's validity, which can be categorized into four dimensions: internal validity, external validity, construct validity, and statistical conclusion validity.

\subsubsection{Internal validity} 
A primary threat to internal validity arises from the variation in sample sizes across the different validation experiments. For example, Experiment 1-MSc included only 14 participants, whereas Experiments 2-BScMay and 4-BScNov involved 84 and 109 subjects, respectively. This disparity may affect statistical power and limit the ability to detect differences attributable solely to circuit complexity. Additionally, some residual biases may have influenced participant performance despite the measures taken to mitigate the impact of contextual variables—such as motivation, fatigue, or learning effects. It is also important to note that Experiment 3-Prof was conducted online without direct supervision, which could have increased data variability due to uncontrolled environmental conditions. Nevertheless, the commitment and expertise of this group of quantum computing and software quality professionals helped mitigate these effects, although the online modality remains a potential threat to internal validity.

\subsubsection{External validity}
Another significant limitation is the generalization of results to other contexts. Most participants were drawn from academic environments, which may introduce bias into the results, as these evaluators typically have a strong educational background and prior experience with analysis and debugging methods. This limits the extrapolation of findings to industrial contexts, where user profiles and working conditions may differ significantly. Therefore, future research should include more heterogeneous samples covering various industrial sectors to strengthen the external validity of the model.

\subsubsection{Construct validity}
Construct validity is related to the ability of the selected metrics to accurately capture the concept of analyzability in quantum systems. Our model integrates classical indicators based on the ISO/IEC 25010 standard with quantum-specific metrics, such as circuit depth and quantum cyclomatic complexity. However, it is possible that these indicators do not fully reflect the intrinsic complexity of the circuits, particularly under conditions of high variability. Additionally, external factors—such as prior experience in classical and quantum computing or the order in which circuits are presented—may complicate the correlation between objective metrics and the subjective perception of complexity, thus affecting construct validity. It is also important to note that most developers and evaluators involved in this study come from academic environments, which may limit the representativeness of the metrics compared to industrial settings, where software development challenges can be significantly different.

\subsubsection{Conclusion validity}
The use of non-parametric statistical methods, such as the Kruskal-Wallis test and the Friedman test, as well as the combination of \(p\)-values using Stouffer's method, is appropriate given that the data do not follow a normal distribution and the sample sizes vary considerably across experiments. However, these techniques assume the independence of observations and a certain level of homogeneity in variance across groups. Any violation of these assumptions could affect the accuracy of inferences. Furthermore, the reliance on a single dataset for analysis limits the validity of the conclusions, highlighting the need to replicate the study using multiple datasets from diverse domains to confirm the consistency of the findings.

\section{Conclusions and Future Work}
\label{conclusions}

The results obtained throughout the family of experiments allow us to draw several conclusions that reinforce the validity and usefulness of the quantum computing component within the proposed analyzability model for hybrid software. The findings from the analysis and meta-analysis reveal that participants' scores differ significantly depending on the level of analyzability of quantum circuits as defined by the model. This evidence empirically supports the model's ability to distinguish between circuits with different levels of structural clarity, showing that higher analyzability correlates with higher scores and less dispersed distributions. This finding is particularly relevant given the diversity of the participant groups, which included students from different academic levels and subjects experienced in quantum computing and software quality, thus strengthening the model's internal and external validity.

Beyond its applicability in academic settings, the model presents significant potential for professional contexts, where the convergence of classical and quantum components offers advantages in optimizing, inspecting, and maintaining complex systems. The ability to effectively differentiate between circuits with higher and lower analyzability enables development teams and quality assurance specialists to prioritize refactoring efforts, identify complexity \textit{hotspots}, and guide the training of new professionals in areas where quantum software poses more significant challenges.

It is important to note that the empirical validation presented in this study has primarily focused on quantum metrics, as classical metrics have already been extensively validated through standards such as ISO/IEC 25010. Therefore, a crucial future research direction will be validating the hybrid model as a whole, effectively integrating both types of metrics. This will involve conducting studies in industrial scenarios and more diverse contexts to confirm the validity and applicability of the hybrid model in real-world environments.

Several initiatives will be undertaken to consolidate and disseminate the benefits of this model. First, efforts will be intensified to promote awareness of the model within both academic and industrial communities, ensuring that its foundations and potential applications are well understood. This will enhance its adoption and facilitate its incorporation into various development environments. In this way, we aim to foster the integration of the model into continuous improvement processes, encouraging collaboration with research teams and institutions interested in advancing quantum software quality. Finally, validation in even broader contexts—such as companies already using or experimenting with quantum technologies—would provide additional insights into the model’s effectiveness and industry reception, contributing to the maturation of quantum software engineering and the advancement of best practices in this emerging field.

\section*{Declarations}

\subsection*{Funding}
This research was supported by the project ATHENA-HYGIEIA (PID2024-155693NB-C42), funded by MCIU/AEI/10.13039/501100011033 / FEDER, UE; the QSERV: Quantum Service Engineering: Development Quality, Testing \& Security of Quantum Microservices (PID2021-124054OB-C32) project funded by the Spanish Ministry of Science and Innovation and ERDF; the Q2SM: Quality Quantum Software Model (13/22/IN/032) project financed by the Junta de Comunidades de Castilla-La Mancha and FEDER funds; the AETHER-UCLM: A holistic approach to smart data for context-guided data analysis (PID2020-112540RB-C42) funded by MCIN/AEI/10.13039/501100011033/; and the project “QUASAR: QUAntum software engineering for Secure, Affordable, and Reliable systems” grant 2022T2E39C under the PRIN 2022 MUR program funded by the EU - NGEU. It also received financial support for the execution of applied research projects within the SINERGIA (2025-GRIN-38310) framework, part of the UCLM Own Research Plan, 85\% co-funded by the European Regional Development Fund (FEDER).

\subsection*{Ethical approval}
Not applicable.

\subsection*{Informed consent}
Not applicable.

\subsection*{Author Contributions}
Ana Díaz Muñoz led the conceptualization, methodology, experimental design, and writing of the manuscript. Maria Teresa Baldassarre contributed to the validation strategy, structure and execution of the study, and review and improvement of the manuscript. José Antonio Cruz-Lemus, Moisés Rodríguez and Mario Piattini supervised the research design, contributed to the validation strategy and to the revision and critical improvement of the manuscript. All authors read and approved the final manuscript.

\subsection*{Data Availability Statement}
\label{availability}
All experimental data, response collection forms, and the source code used for statistical analysis are publicly available to the community. These materials can be found in an online repository\footnote{\url{https://github.com/aniitadiazm/EmSE2025-Diazetal}}, enabling the replication and extension of the experiments presented in this article.

The repository is structured into several directories, each serving a specific purpose:

\begin{itemize}
    \item \textbf{datasets/}: Contains the anonymous response data collected from all participants across the four experiments.
    \item \textbf{questionnaires/}: Includes all the forms used for circuit comprehension evaluation, with examples of the types of questions posed.
    \item \textbf{study\_materials/}: Provides instructional resources and worked examples to guide participants during the tasks.
    \item \textbf{experimental\_tasks/}: Describes the circuits, instructions, and task flow used in the experiments.
    \item \textbf{support\_materials/}: Contains supplementary resources offered during tutoring sessions or for optional reinforcement.
\end{itemize}

This open-access repository facilitates transparency, reproducibility, and further research in the field of quantum software quality evaluation.

\subsection*{Conflict of Interest}
The authors declare that they have no conflict of interest.

\subsection*{Clinical Trial Number}
Clinical trial number: not applicable.

\bibliographystyle{spmpsci}
\bibliography{references}

\end{document}